\documentclass[a4paper,11pt]{article}
\pdfoutput=1

\usepackage{jcappub}

\usepackage[T1]{fontenc}

\usepackage{aas_macros}

\title{\boldmath Scale-dependent dipolar modulation and the
  quadrupole-octopole alignment in the CMB temperature}

\author{A. Marcos-Caballero}
\author{and E. Mart\'inez-Gonz\'alez}
\affiliation{Instituto de F\'isica de Cantabria, CSIC-Universidad de
  Cantabria,\\Avda. de los Castros s/n, 39005 Santander, Spain.}

\emailAdd{marcos@ifca.unican.es}
\emailAdd{martinez@ifca.unican.es}

\abstract{The connection between the dipolar modulation asymmetry and the
quadrupole-octopole alignment in the CMB is studied in this
work. First, a generalization of the dipolar modulation model is
proposed by considering that the amplitude may depend on the scale. As
derived from a Bayesian inference analysis, this model fits the CMB
data better than the scale-independent one. As an extension of the
standard model, the scale-dependent dipolar modulation shows
comparable evidence to the standard isotropic model in the large
scales ($\ell_\mathrm{max} \leq 64$). The posterior distribution of
the parameters of the scale-dependent model suggests that the
amplitude of the dipolar modulation is large at the lowest
multipoles. This large asymmetry induces a detectable correlation
between the quadrupole and the octopole. The significance of the
quadrupole-octopole alignment is analyzed under the assumption that
the Universe has a scale-dependent dipolar modulation. The three
alignment estimators considered in this paper show an increment of
$80\%$ in the p-value, showing a clear correlation between these two
CMB anomalies. Within this new scenario, only one of the alignment
estimators is still below the $1\%$ probability level.}

\begin{document}
\maketitle
\flushbottom

\section{Introduction}

The standard cosmological model provides a good fit to the recent
precise observations of the temperature and polarization of the Cosmic
Microwave Background (CMB) radiation \cite{planck012018}. However,
there exists some slight deviations in the temperature field at the
largest scales which might indicate that new physics is needed in
order to have a full description of the Universe. Several anomalies
have been reported in the literature, many of them reflecting that the
large-scale CMB anisotropies do not follow the statistical isotropy
expected in the standard model. In particular, CMB data reflects that
there is an hemisphere of the sky which has more power than the
opposite one
\cite{eriksen2004,hansen2009,akrami2014,planck232013,planck162015,planck072018}. Different
estimators have been proposed to characterize this anomaly. In
particular, the dipolar modulation model allow us to determine what is
the preferred direction on the sky and the degree of asymmetry
observed in the CMB temperature \cite{hoftuft2009}. It is plausible
that some of these deviations are not independent (see \cite{muir2018}
for an analysis of the correlation among the most common large-scale
isotropy estimators within the standard model).

Another well-known large-scale anomaly is the alignment between the
quadrupole and the octopole \cite{deOliveira-Costa2004,copi2015}. This
alignment clearly indicates the existence of a preferred direction in
the CMB temperature. Indeed, the alignment of the lowest multipoles
could be responsible for the large-scale structures observed in the
Ecliptic southern hemisphere \cite{copi2006,marcos-caballero2017}. The
coincidence that these structures are located in the same region of
the sky where the variance asymmetry is observed suggests an eventual
relation between the dipolar modulation and the quadrupole-octopole
alignment. This possible correlation between the alignment and the
(scale-invariant) dipolar modulation has been already studied in
\cite{gordon2007,hoftuft2009}. It was found a negligible connection
between both anomalies in this case. In this work, we propose to
generalize the standard dipolar modulation model by including a scale
dependence in the amplitude. Evidences that the dipolar modulation may
depend on the scale have been already observed in \cite{planck162015}
by computing the modulation on different multipole intervals. In
particular, the amplitude of the modulation at the largest scales
seems to be greater than the ones derived from smaller scales. This
may cause that the effective modulations of the quadrupole and
octopole moments are larger than expected from the standard
scale-invariant model, and hence, a non-negligible effect on the
quadrupole-octopole alignment may be observed.

This paper is organized as follows: the scale-dependent dipolar
modulation model is introduced in section~\ref{sec:dm_model} as a
generalization of the scale-invariant one. The inpainting procedure
used in the analysis of the CMB temperature is described in
section~\ref{sec:inpainting}, whereas the iterative posterior
estimation method for these maps and the details of the likelihood
calculation are shown in sections~\ref{sec:posterior_estimation} and
\ref{sec:likelihood}. The results on the scale-dependent dipolar
modulation model are presented in
section~\ref{sec:results_dm}. Finally, the quadrupole-octopole
alignment estimators and the corresponding results derived from them
for the scale-dependent dipolar modulation model are shown in
sections~\ref{sec:qo_alignment} and \ref{sec:results_qo},
respectively. The overall conclusions of the paper are presented in
section~\ref{sec:conclusions}.

\section{Dipolar modulation model}
\label{sec:dm_model}

A non-isotropic model of the CMB temperature based on the modulation
of an isotropic field has been proposed to reproduce the hemispherical
asymmetry observed in the CMB \cite{gordon2007,hoftuft2009}. If the
dipolar model is tested at different scales, then a clear dependence
with the multipole is observed \cite{planck162015}. In this section,
we introduce the scale-dependent dipolar modulation model by
generalizing the scale-independent one previously considered in the
literature.

\subsection{Scale-independent modulation}

In general, anisotropic models of the CMB temperature can be obtained
by multiplying an isotropic field $T$ by a modulating function:
\begin{equation}
\hat{T}(\mathbf{n}) = M(\mathbf{n}) \ T(\mathbf{n}) \ ,
\label{eqn:modulation}
\end{equation}
where $M$ can be expanded in terms of the spherical harmonics:
\begin{equation}
M(\mathbf{n}) = \sum_{\ell=}^\infty \sum_{m=-\ell}^\ell M_{lm}
\ Y_{\ell m} (\mathbf{n}) \ .
\end{equation}
In the case of the dipolar modulation, we assume that the modulating
function only has contributions from the monopole ($\ell=0$) and the
dipole ($\ell=1$). Therefore, it is considered that
\begin{equation}
M(\mathbf{n}) = 1 + \mathbf{A} \cdot \mathbf{n} \ ,
\end{equation}
where $\mathbf{A}$ is a three-dimensional vector parameterizing the
dipolar modulation. Notice that the monopole ($M_{00}$) is chosen such
that the isotropic model is recovered when $\mathbf{A} =
0$. Extensions of this model could include a monopole term in order to
account for an isotropic low variance.

\subsection{Scale-dependent modulation}

It is possible to generalize the modulation by considering a scale
dependence in the model. First, we decompose the isotropic CMB
temperature field in the different multipole moments:
\begin{equation}
T(\mathbf{n}) = \sum_{\ell=0}^\infty T_\ell(\mathbf{n}) \ ,
\end{equation}
where the field $T_\ell$ represents the $\ell$-th multipolar moment of
the field. These moments can be expanded in terms of the spherical
harmonics:
\begin{equation}
T_\ell(\mathbf{n}) = \sum_{m=-\ell}^\ell a_{\ell m} Y_{\ell
  m}(\mathbf{n}) \ .
\end{equation}
In this work, we consider a generalization of the model presented in
eq.~\eqref{eqn:modulation} in the following way:
\begin{equation}
\hat{T}(\mathbf{n}) = \sum_{\ell=0}^\infty M_\ell(\mathbf{n})
\ T_\ell(\mathbf{n}) \ ,
\label{eqn:scale-dependent_modulation}
\end{equation}
where the modulating function depends on the multipole. In particular,
we have the following expression for the scale-dependent dipolar
modulation:
\begin{equation}
M_\ell(\mathbf{n}) = 1 + \mathbf{A}_\ell \cdot \mathbf{n} \ .
\label{eqn:Mell}
\end{equation}
In general, the parameter space of this model is given by the
components of the vectors $\mathbf{A}_\ell$. In order to simplify the
model and reduce the number of parameters, we consider the following
parametric expression for these vectors:
\begin{equation}
\mathbf{A}_\ell = \mathbf{A} \left( \frac{\ell_0}{\ell} \right)^\alpha
\ .
\label{eqn:Al}
\end{equation}
We have assumed that the direction of the dipolar modulation is the
same for all the scales, which is given by the amplitude vector
$\mathbf{A}$. However, the amplitude depends on multipole following a
power-law with tilt $\alpha$. This model of scale-dependent
dipolar-modulation has four parameters: the three components of the
vector $\mathbf{A}$, which characterizes the direction and the
amplitude of the modulation at the pivot multipole $\ell_0$, and the
parameter $\alpha$ representing the scale dependence. The pivot
multipole $\ell_0$ can be freely chosen, since it is not a parameter
of the model (we can change the pivot multipole by redefining the
amplitude of the vector $\mathbf{A}$). However, it is possible to
reduce the correlation between $\mathbf{A}$ and $\alpha$ by
considering a suitable pivot scale. For this reason, we use the value
$\ell_0=5$ for the pivot multipole in our analysis. Finally, notice
that the scale-independent model is recovered when $\alpha=0$.

It is possible to write the scale-dependent model given by
eqs.~\eqref{eqn:scale-dependent_modulation} and \eqref{eqn:Mell} as a
convolution is real space:
\begin{equation}
\hat{T}(\mathbf{n}) = \left[ 1 + \mathbf{n} \cdot \mathbf{A}(\mathbf{n}) \otimes
\right] T(\mathbf{n}) \ ,
\label{eqn:dm_model}
\end{equation}
where the function $\mathbf{A}(\mathbf{n})$ is given by the filter
coefficients $\mathbf{A}_\ell$.

\begin{figure}
\begin{center}
\includegraphics[width=0.49\textwidth]{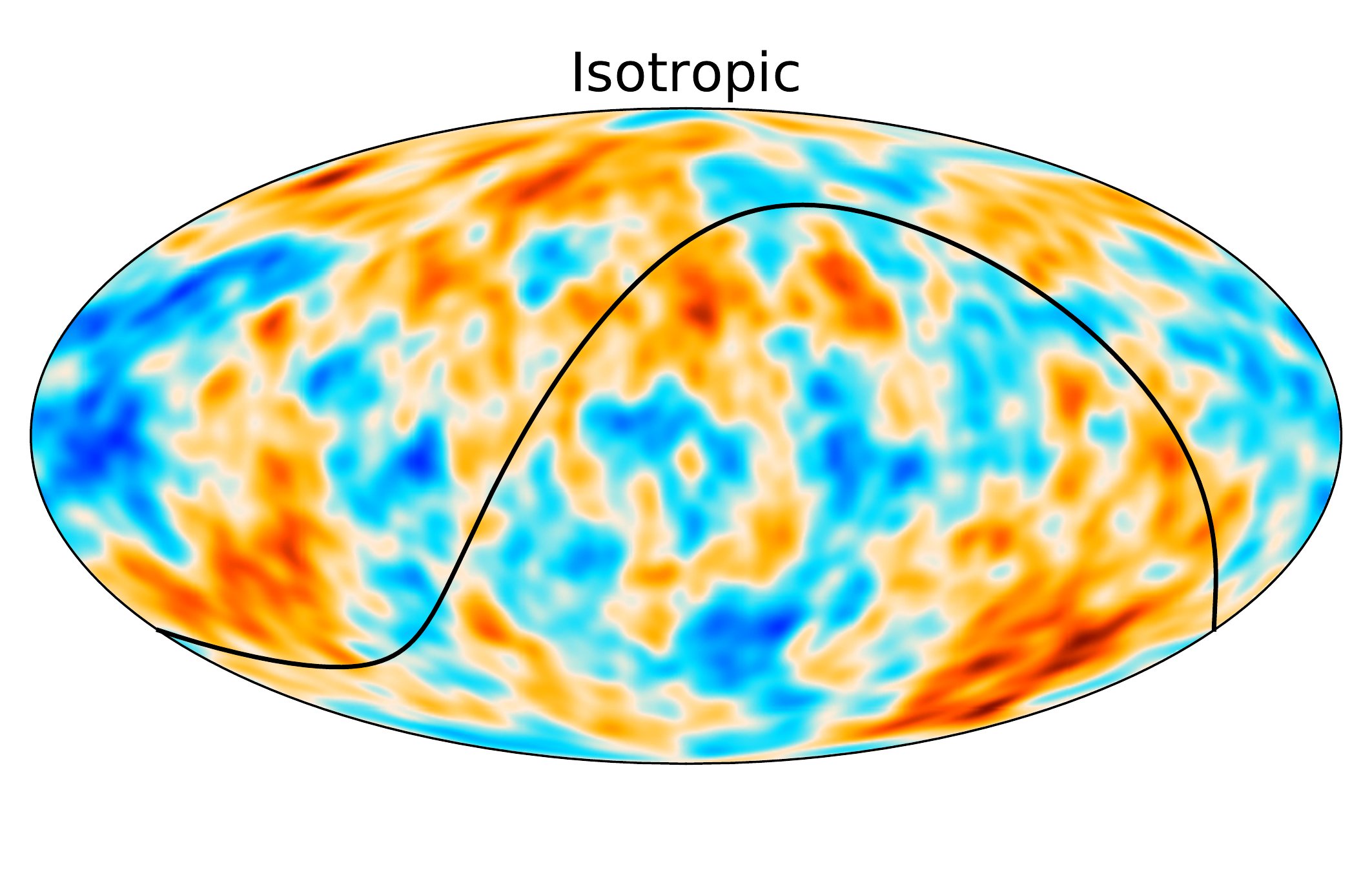}
\end{center}
\begin{center}
\includegraphics[width=0.49\textwidth]{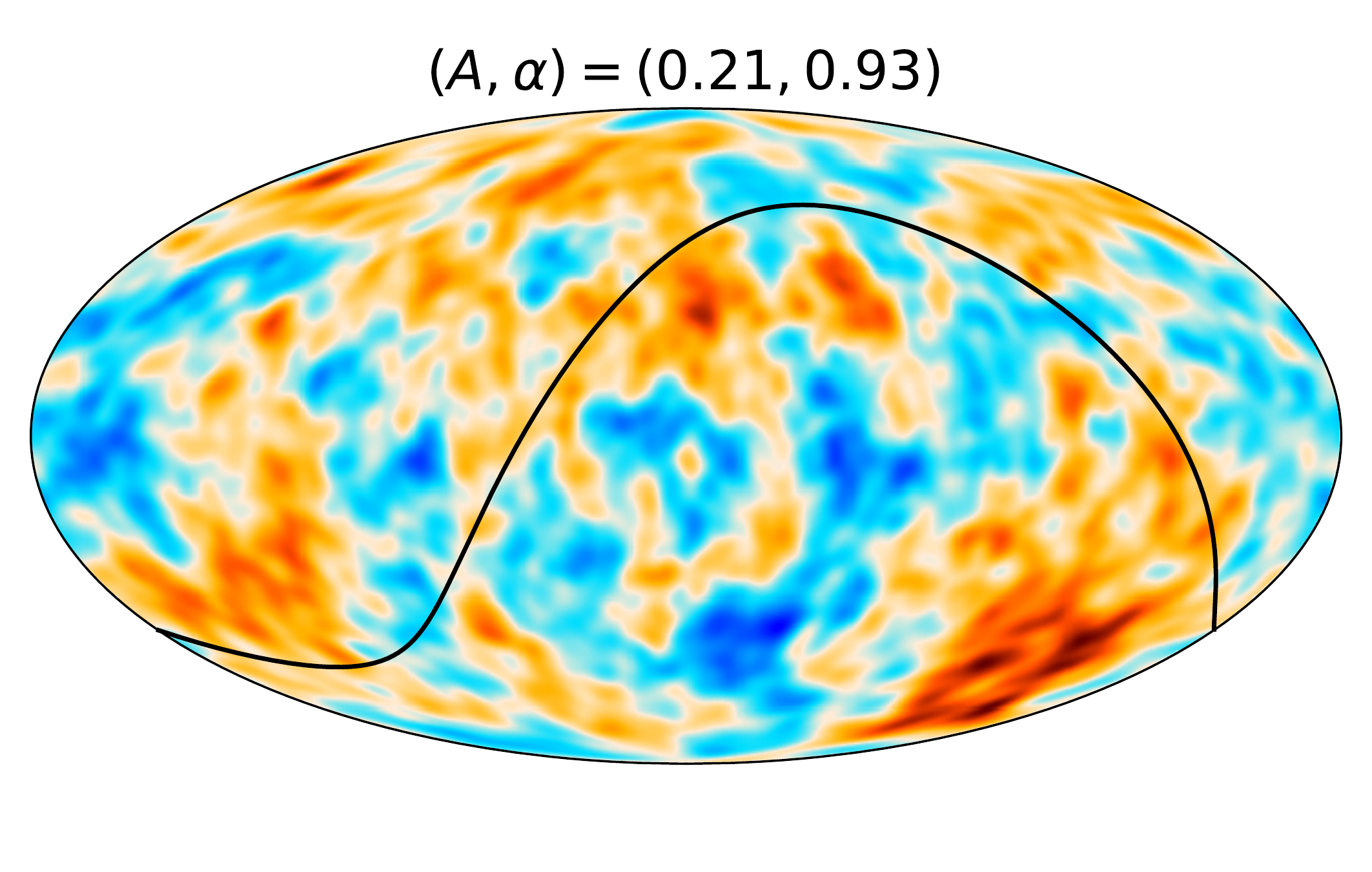}
\includegraphics[width=0.49\textwidth]{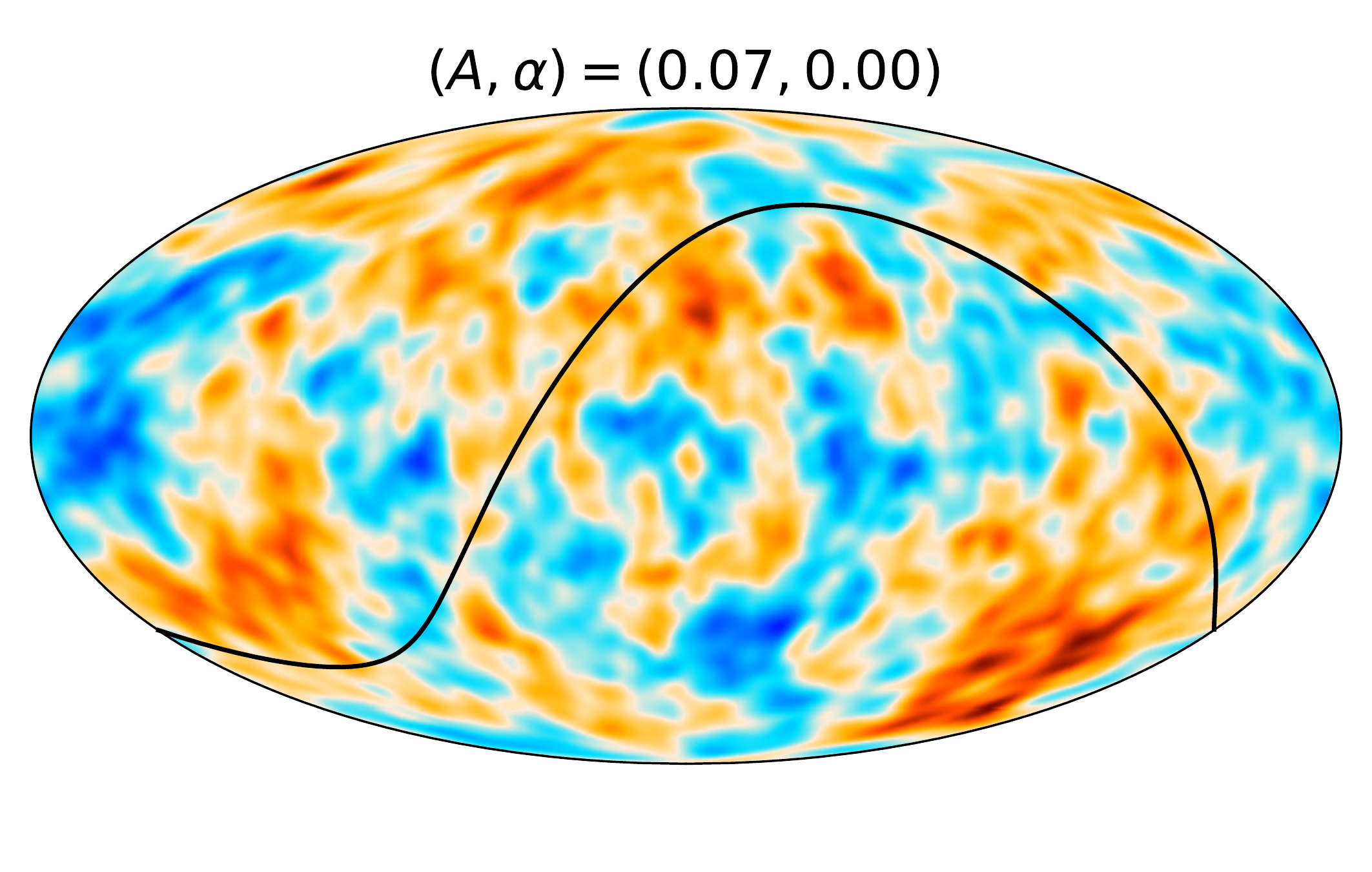}
\includegraphics[width=0.49\textwidth]{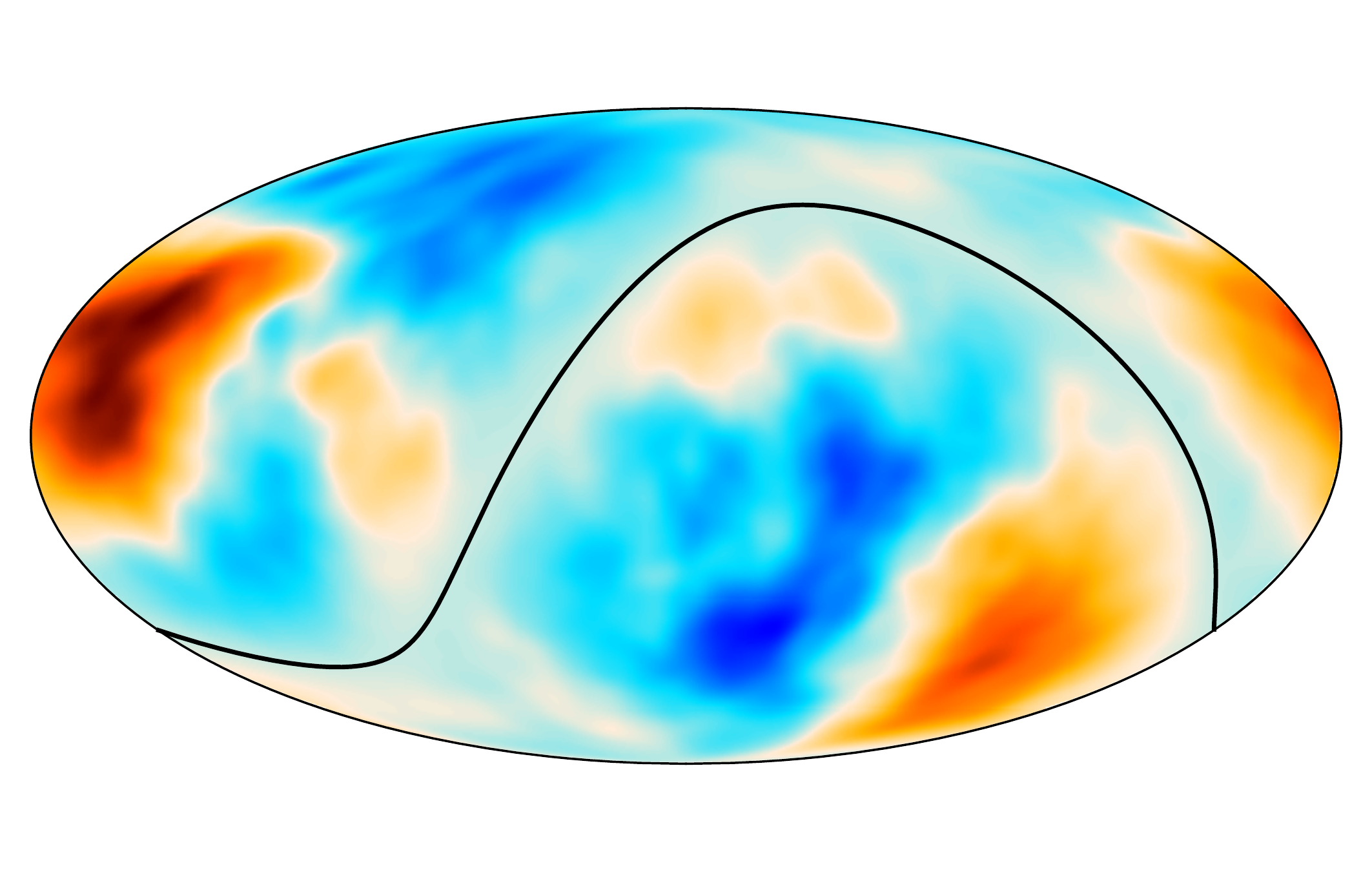}
\includegraphics[width=0.49\textwidth]{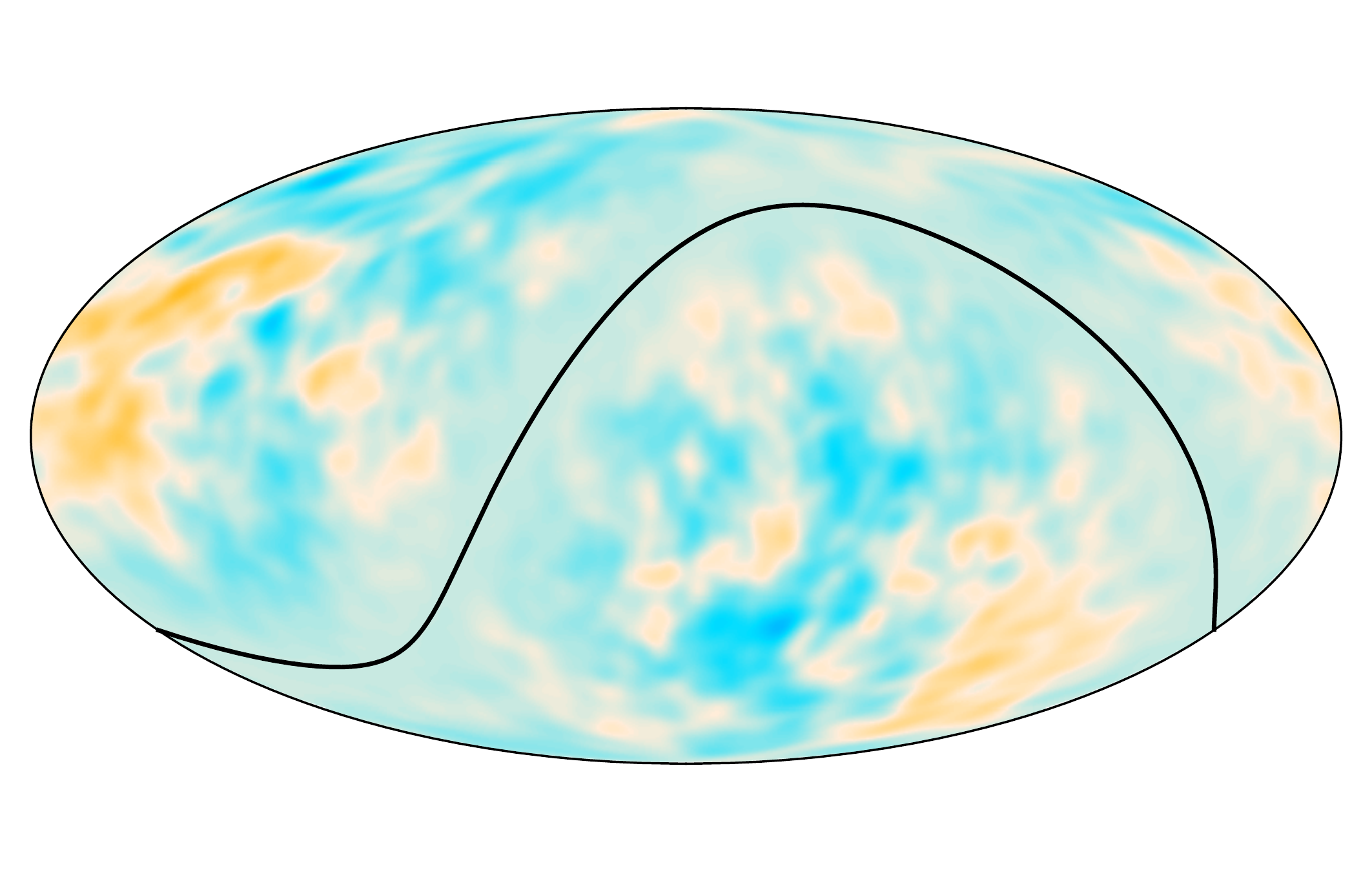}
\end{center}
\caption{Comparison between the scale-dependent and scale-invariant
  dipolar modulation models. All the simulations in this figure have a
  FWHM of $5^\circ$. The equator of the dipolar modulation considered
  in these maps is represented by the black lines. \emph{Upper map:}
  CMB simulation according to the isotropic $\Lambda$CDM
  model. \emph{Middle maps:} the scale-dependent and the
  scale-invariant dipolar modulation models calculated by modulating
  the previous isotropic simulation. The parameters of the model are
  $(A,\alpha) = (0.21, 0.93)$ for model with the scale-dependent
  amplitude (left) and $(A,\alpha) = (0.07,0.)$ for the standard
  dipolar modulation with constant amplitude (right). The dipolar
  modulation amplitudes are referred to the multipole
  $\ell=5$. \emph{Bottom maps:} differences of the dipolar modulation
  models shown above with respect to the isotropic model.}
\label{fig:sim_dm}
\end{figure}

In figure~\ref{fig:sim_dm}, it is shown a simulation of the dipolar
modulation considering both the scale-dependent and the
scale-invariant models described above. The amplitude of the
scale-invariant model is given in \cite{hoftuft2009}, $A=0.07$ and the
parameters of the scale-dependent model are chosen to be similar to
the ones derived in this work. The effect of the dipolar modulation on
the lower multipoles is enhanced in the scale-dependent model with
respect to the case of the standard dipolar modulation with constant
amplitude. In particular, this may imply a larger correlation between
the low multipoles.

\section{Inpainting with constrained realizations}
\label{sec:inpainting}

Since the CMB temperature maps may have foregrounds residuals, a
confidence mask is provided by the component separation methods
covering mainly the galactic plane and other regions with high
foreground contamination. The incomplete sky makes that spherical
harmonic analyses are complicated due to multipole correlations and
the presence of a bias in the angular power spectrum
estimation. However, this methodological difficulties can be overcome
if the mask region is filled with inpainted data.

Assuming that we have a probabilistic model for our sample, the
inpainting can be done coherently with this model and the available
data outside the mask. The procedure of inpainting with constrained
realizations relies on the calculation of the conditional probability
density $p(\hat{\mathbf{d}}|\mathbf{d})$, where $\hat{\mathbf{d}}$ is
a vector representing the inpainted field and $\mathbf{d}$ corresponds
to the data we are able to observe. Under the hypothesis that the
field is Gaussian, we only need the pixel covariance matrix, which can
be calculated from the angular power spectrum $C_\ell$:
\begin{equation}
\mathbf{C}_{ij} = \sum_{\ell=0}^{\ell_\mathrm{max}} \frac{2\ell
  +1}{4\pi} \ C_\ell \ P_\ell(\mathbf{n}_i \cdot \mathbf{n}_j) \ ,
\end{equation}
where $\mathbf{n}_i$ is the unit vector representing the $i$th-pixel.
The maximum multipole $\ell_\mathrm{max}$ included in the sum is
considered large enough to ensure that the covariance matrix is not
singular and positive definite. In our analysis, the value
$\ell_\mathrm{max} = 160$ is used for maps with $N_\mathrm{side} =
32$. Since the matrix $\mathbf{C}$ has dimension $N_\mathrm{pix}
\times N_\mathrm{pix}$, we are limited to low resolution maps due to
numerical capabilities.

The procedure we follow to calculate the constrained realizations does
not need any matrix inversion. The first step consists in reordering
the columns and rows of $\mathbf{C}$ such that all the observed pixels
are in the first entries and the masked pixels in the last
indices. Once we have this particular order in the new matrix
$\bar{\mathbf{C}}$, its Cholesky decomposition is calculated:
\begin{equation}
\mathbf{C} = \bar{\mathbf{L}} \bar{\mathbf{L}}^t \ ,
\end{equation}
where $\bar{\mathbf{L}}$ is a lower triangular matrix, which has the
following representation in blocks:
\begin{equation}
\bar{\mathbf{L}} = \left( \begin{array}{cc}
\mathbf{L} & 0 \\
\mathbf{R} & \hat{\mathbf{L}} \\
\end{array}\right) \ .
\end{equation}
Whilst $\mathbf{L}$ and $\hat{\mathbf{L}}$ in this expression are both
lower triangular matrices, $\mathbf{R}$ is a rectangular matrix with
dimensions the number of masked pixels by the number of observed
pixels.

The Cholesky decomposition allow us to calculate Gaussian realizations
of the field by calculating the product
\begin{equation}
\left( \begin{array}{c}
\mathbf{d} \\ \hat{\mathbf{d}}
\end{array} \right) = 
\left( \begin{array}{cc}
\mathbf{L} & 0 \\
\mathbf{R} & \hat{\mathbf{L}} \\
\end{array}\right)
\left( \begin{array}{c}
\mathbf{z} \\ \hat{\mathbf{z}}
\end{array} \right) \ ,
\end{equation}
where $\mathbf{z}$ and $\hat{\mathbf{z}}$ are Gaussian random numbers
with zero mean and unit variance. Since we are interested in
generating the field $\hat{\mathbf{d}}$ given the data $\mathbf{d}$,
the following system of equations has to be solved
\begin{equation}
\mathbf{d} = \mathbf{L} \mathbf{z} \ ,
\end{equation}
\begin{equation}
\hat{\mathbf{d}} = \mathbf{R}\mathbf{z} + \hat{\mathbf{L}} \hat{\mathbf{z}} \ .
\end{equation}
The first linear equation can be solve by recursive substitution to
obtain the vector $\mathbf{z}$. This solution is denoted by
$\mathbf{L}^{-1} \mathbf{d}$, although we do not need to perform the
matrix inversion explicitly in the numerical implementation. Finally,
the inpainted data $\hat{\mathbf{d}}$ is given by
\begin{equation}
\hat{\mathbf{d}} = \mathbf{R} (\mathbf{L}^{-1} \mathbf{d}) +
\hat{\mathbf{L}} \hat{\mathbf{z}} \ .
\end{equation}
From this expression, it is possible to deduce that $\hat{\mathbf{d}}$
is Gaussian random field with the following mean and covariance:
\begin{equation}
\langle \hat{\mathbf{d}} \rangle = \mathbf{R} (\mathbf{L}^{-1}
\mathbf{d}) \ ,
\label{eqn:constrained_realizations_mean}
\end{equation}
\begin{equation}
\hat{\mathbf{C}} = \langle \hat{\mathbf{d}} \hat{\mathbf{d}}^t \rangle
- \langle \hat{\mathbf{d}} \rangle \langle \hat{\mathbf{d}}^t \rangle
= \hat{\mathbf{L}} \hat{\mathbf{L}}^t \ .
\label{eqn:constrained_realizations_cov}
\end{equation}
Notice that the values of the observed pixels only affects to the mean
of $\hat{\mathbf{d}}$, not to its covariance. However, the particular
geometry of the mask and the theoretical model used to reconstruct the
masked pixels have an influence in both the mean and the covariance of
$\hat{\mathbf{d}}$.

\section{Iterative posterior estimation}
\label{sec:posterior_estimation}

In this section, we introduce a novel method to estimate the posterior
distributions which are given by inpainted data. Sometimes, it is
difficult to calculate the likelihood function for incomplete data
when the procedure involves convolutions or Fourier transforms. The
simulation of inpainted data is useful to avoid this problems.
However, it is important to take into account that fictitious data are
included in the analysis. Moreover, the inpainted region could bias
the results because they are generated according to a given fiducial
model, which may be different to the one preferred by the available
data. In the case of parameter estimation, the information of the
inpainted data can be removed from the posterior distribution by
applying the iterative method explained below.

The posterior distribution of the model parameters $\theta$ is given
by
\begin{equation}
p(\theta|\mathbf{d}, \hat{\mathbf{d}}) = \frac{\pi(\theta)
  \ L(\mathbf{d}, \hat{\mathbf{d}}|\theta)}{Z(\mathbf{d},
  \hat{\mathbf{d}})} \ ,
\end{equation}
where $\mathbf{d}$ and $\hat{\mathbf{d}}$ are the available data and
inpainted data, respectively. In this equation, $\pi$ represents the
prior distribution, $L$ the likelihood function and $Z$ the evidence
of the model. We can calculate the posterior distribution of the
parameters given only the observed data $\mathbf{d}$ by averaging the
posterior $p(\theta|\mathbf{d}, \hat{\mathbf{d}})$ over the
$\hat{\mathbf{d}}$ vector:
\begin{equation}
p(\theta|\mathbf{d}) = \int p(\theta|\mathbf{d}, \hat{\mathbf{d}})
\ p(\hat{\mathbf{d}}|\mathbf{d}) \ \mathrm{d}\hat{d} \ ,
\label{eqn:posterior_integral}
\end{equation}
where $p(\hat{\mathbf{d}}|\mathbf{d})$ is the probability of
$\hat{\mathbf{d}}$ given the data $\mathbf{d}$. In practice, the
probability distribution $p(\hat{\mathbf{d}}|\mathbf{d})$ cannot be
calculated because we do not know the underlying model:
\begin{equation}
p(\hat{\mathbf{d}}|\mathbf{d}) = \int p(\hat{\mathbf{d}}|\mathbf{d},
\theta) \ p(\theta|\mathbf{d}) \ \mathrm{d}\theta \  .
\label{eqn:prob_data_integral}
\end{equation}
The probability density $p(\hat{\mathbf{d}}|\mathbf{d},\theta)$ is the
one used to calculate the constrained realizations, which in our case
is a Gaussian distribution with mean and variance given in
eqs.~\eqref{eqn:constrained_realizations_mean} and
\eqref{eqn:constrained_realizations_cov}, respectively. However, we
need the posterior density $p(\theta|\mathbf{d})$ to calculate this
integral. For this reason, an iterative method is proposed to find the
posterior. This method is based on the calculation of a sequence of
posterior densities $p_n(\theta|\mathbf{d})$ such that they converge
to the real posterior $p(\theta|\mathbf{d})$. The recursive equations
are given by the expressions in eqs.~\eqref{eqn:posterior_integral}
and \eqref{eqn:prob_data_integral}. The $n$-th posterior
$p_{n}(\theta|\mathbf{d})$ is calculated from the following integral
\begin{equation}
p_{n}(\theta|\mathbf{d}) = \int p(\theta|\mathbf{d}, \hat{\mathbf{d}})
\ p_{n}(\hat{\mathbf{d}}|\mathbf{d}) \ \mathrm{d}\hat{d} \ ,
\label{eqn:posterior_integral_ite}
\end{equation}
where $p_{n}(\hat{\mathbf{d}}|\mathbf{d})$ depends on the posterior in
the previous step:
\begin{equation}
p_{n}(\hat{\mathbf{d}}|\mathbf{d}) = \int
p(\hat{\mathbf{d}}|\mathbf{d}, \theta) \ p_{n-1}(\theta|\mathbf{d})
\ \mathrm{d}\theta \ .
\label{eqn:prob_data_integral_ite}
\end{equation}

We summarize the iterative procedure in the following steps, assuming
that a Markov Chain Monte Carlo (MCMC) method is used for the
sampling:
\begin{enumerate}
\item We choose a initial posterior $p_{0}(\theta|\mathbf{d})$, which
  must be close to the real posterior in order to have a fast
  convergence. The initial guess can be a Dirac delta at the fiducial
  model $\theta_0$, typically the best-fit $\Lambda$CDM model.
\item We simulate a set of constrained realizations by using the
  probability $p(\hat{\mathbf{d}}|\mathbf{d}, \theta)$, which depends
  on the observed data $\mathbf{d}$ and the parameters of the
  model. In this case, the set of parameters $\theta$ assumed in each
  constrained realization are sampled with the distribution
  $p_{n-1}(\theta|\mathbf{d})$. For instance, the parameters are fixed
  to the fiducial model in all the constrained realizations in our
  initial guess ($n=1$). At the end of this step, we have a set of $N$
  inpainted data $(\mathbf{d},\hat{\mathbf{d}}_i)$ ($i = 1,\dots, N$),
  where the probability distribution of the vectors
  $\hat{\mathbf{d}}_i$ is
  $p_{n}(\hat{\mathbf{d}}_i|\mathbf{d})$. Notice that we do not need
  to calculate the integral in eq.~\eqref{eqn:prob_data_integral_ite}
  explicitly in our methodology. Only samples of inpainted data
  following the conditional distribution
  $p_{n}(\hat{\mathbf{d}}|\mathbf{d})$ are needed.

\item We estimate the posterior for each data set
  $(\mathbf{d},\hat{\mathbf{d}}_i)$ (typically by running a MCMC
  sampler), obtaining the set of posteriors $p(\theta|\mathbf{d},
  \hat{\mathbf{d}}_i)$.
\item The posteriors $p(\theta|\mathbf{d}, \hat{\mathbf{d}}_i)$ are
  averaged in order to obtain the next posterior in the iteration
  procedure:
\begin{equation}
  p_{n+1}(\theta|\mathbf{d}) = \frac{1}{N} \sum_{i=1}^N
  p(\theta|\mathbf{d}, \hat{\mathbf{d}}_i) \ .
\end{equation}
The result of this expression is a Monte Carlo estimate of the
integral in eq.~\eqref{eqn:posterior_integral_ite}. If the sampling
method is based on a MCMC algorithm, we can calculate posterior
samples directly from the Markov chains. Once the burn-in points are
removed from the chains, a single set of samples of the parameter
space is obtained for each posterior $p(\theta|\mathbf{d},
\hat{\mathbf{d}}_i)$ by combining all the samples of the chains
generated by the MCMC algorithm. Finally, the union of all the samples
of the posteriors $p(\theta|\mathbf{d}, \hat{\mathbf{d}}_i)$
($i=1,\dots,N$) characterizes the distribution of the posterior
$p_{n}(\theta|\mathbf{d})$\footnote{Here we have assumed that the
  number of samples of each posterior $p(\theta|\mathbf{d},
  \hat{\mathbf{d}}_i)$ are the same. If not, we have to weight the
  samples in order to assure that all the posteriors
  $p(\theta|\mathbf{d}, \hat{\mathbf{d}}_i)$ contribute equally to the
  mean posterior $p_{n}(\theta|\mathbf{d})$.}.
\item The posterior $p_{n}(\theta|\mathbf{d})$ obtained in the
  previous step is used in step 2 (as the posterior
  $p_{n}(\theta|\mathbf{d})$) to generate the new set of inpainted
  data for the next iteration. Notice that the set of parameters
  $\theta$ needed to calculated the constrained realizations in the
  step 2 can be obtained directly by taking random samples of the
  posterior $p_{n+1}(\theta|\mathbf{d})$ as calculated in the step 4.
\end{enumerate}

\section{Likelihood}
\label{sec:likelihood}

Since we analyze full-sky constrained realizations in the iterative
method explained above, the likelihood in harmonic space is simpler
than in the case of masked data due to the lack of correlations
between different multipoles. In order to calculate the likelihood,
the first step is to relate the harmonic coefficients of the modulated
field $\hat{T}$ with the ones characterizing the isotropic field
$T$. These equations have been obtained for the scale-independent case
in previous analysis of the dipolar modulation \cite{moss2011}. Here
we generalize that expressions for the scale-dependent dipolar
modulation (see appendix~\ref{app:dm_spherical_harmonics}):
\begin{equation}
\hat{a}_{\ell m} = a_{\ell m} + F_{\ell-1 m} A_{\ell-1} a_{\ell-1 m} +
F_{\ell m} A_{\ell+1} a_{\ell+1 m} \ ,
\label{eqn:hat_alm}
\end{equation}
where $A_\ell$ is the modulus of the vector $\mathbf{A}_\ell$ and the
coefficients $F_{\ell m}$ are given by
\begin{equation}
F_{\ell m} = \frac{\ell+1}{\sqrt{(2\ell+1)(2\ell+3)}}
\sqrt{1-\left(\frac{m}{\ell+1}\right)^2} \ .
\label{eqn:Flm}
\end{equation}
In order to obtain these equations, we have assumed the dipolar
modulation directions are the same for all the multipoles (the vectors
$\mathbf{A}_\ell$ are parallel) and that direction is aligned with the
positive $z$ axis. By choosing this particular system of reference on
the sphere, the coefficients with different $m$'s are not couple in
eq.~\eqref{eqn:hat_alm}, which simplifies its inversion. For a given
$m$, eq.~\eqref{eqn:hat_alm} corresponds to a system of linear
equations whose matrix is tridiagonal. The solution of this system can
be obtained very fast by using a simplified version of the Gaussian
elimination algorithm (see appendix~\ref{app:inversion_alm}). During
this process of inversion, the Jacobian of the transformation can also
be calculated as a byproduct.

Given a scale-dependent dipolar modulation model $\theta =
(\mathbf{A}, \alpha)$, the likelihood function is calculated as
follows:
\begin{enumerate}
\item The spherical harmonics transform of the full-sky data $\hat{T}$
  is calculated. Notice that this calculation does not depend on the
  parameters of the model $\theta$, and therefore, it can be done only
  one time at the beginning of the posterior sampling.
\item The spherical harmonics coefficients calculated in the previous
  step are rotated according to the dipolar modulation direction given
  by the unit vector $\mathbf{n}_A \equiv \mathbf{A}/|\mathbf{A}|$, so
  that the direction of asymmetry corresponds to the $z$ axis. The
  resulting coefficients are the $\hat{a}_{\ell m}$ that appear in
  eq.~\eqref{eqn:hat_alm}.
\item The coefficients $a_{\ell m}$ of the associated isotropic field
  are calculated by solving the system of linear equations in
  eq.~\eqref{eqn:hat_alm}. Also the Jacobian $J$ of this
  transformation is obtained in the inversion procedure (see
  appendix~\ref{app:inversion_alm}).
\item Finally, the likelihood function in terms of the coefficients
  $a_{\ell m}$ is given by the Gaussian probability density:
  \begin{equation}
    \ln L(\theta) = - \sum_{\ell=2}^{\ell_\mathrm{max}}
    \frac{\sum_{m=-\ell}^\ell |a_{\ell m}(\theta)|^2}{2 C_\ell
      f_\ell^2} - \ln J(\theta) \ ,
  \end{equation}
  where the Jacobian $J$ is included due to the change of variables
  done before, from $\hat{a}_{\ell m}$ to $a_{\ell m}$. The
  coefficients $f_\ell$ in this equation represent the smoothing
  filters present in our data (typically resolution beam and pixel
  window function).
\end{enumerate}

\section{Results on the scale-dependent dipolar modulation model}
\label{sec:results_dm}

In this section, the results on the scale-dependent dipolar modulation
are given and discussed. In the following subsections, we present the
parameter distribution obtained in the sampling of the posterior, and
the Bayesian inference analysis comparing the scale-dependent model
with the scale-invariant one and the standard model. In addition, a
CMB temperature field without the dipolar modulation is estimated from
the data and the dipolar modulation parameters. A map of the dipolar
modulation anomaly is also given at the end of this section.

The CMB data used in the analysis are the temperature maps provided by
the Planck collaboration \cite{planck012018,planck042018}. We perform
the inpainting on maps as described in section~\ref{sec:inpainting} at
the Healpix resolution $N_\mathrm{side} = 32$. The temperature field
is smoothed with both a Gaussian filter whose FWHM is $2^\circ$ and
the corresponding pixel window function of the resolution. The maximum
multipole considered in the likelihood is $\ell_\mathrm{max} = 64$. We
have verified that the procedure with these specifications does not
have any measurable systematic effect due to the pixelization,
inpainting or any other approximate numerical calculation. In
addition, we have compared the results obtained from two different
component separation methods (SEVEM and SMICA) \cite{planck042018} up
to the first iteration in the posterior estimation method described in
section~\ref{sec:posterior_estimation}. Similar posterior
distributions of the dipolar modulation parameters are recovered from
the two data sets.

\subsection{Parameter estimation}

As explained in the previous sections, the iterative posterior
estimation is applied to inpainted CMB data. The sampling of the
posterior is done by an affine-invariant Markov Chain Monte Carlo
method \cite{goodman2010,emcee2013}. These numerical computations
provide a sample of the posterior once the burn-in period is
removed. The sampled parameters are the three Cartesian components of
the amplitude vector $\mathbf{A}$ at the pivot scale $\ell_0 = 5$ and
the index $\alpha$ (four parameters in total). Once chains for these
parameters are obtained, we can apply the appropriate transformation
in order to infer samples of the amplitude $A$ (the modulus of
$\mathbf{A}$) and dipolar modulation direction $\mathbf{n}_A$.

The methodology used for the parameter estimation has been tested with
a simulated temperature field with scale-dependent dipolar
modulation. All the steps, including the inpainting, have been done
for this simulation. There is not observed bias or systematic effect
on the recovered parameters. Similar posterior than the one obtained
from the real data is obtained for this simulated temperature map.

The prior for the dipolar modulation parameters assumed in the
sampling is given by the natural limitations of the model. In order to
have a positive define covariance matrix, the amplitude of the dipolar
modulation must be less than one for all the multipoles. Otherwise,
the modulating function in eq.~\eqref{eqn:modulation} vanishes for
some directions $\mathbf{n}$ resulting in a singular covariance matrix
for the modulated temperature $\hat{T}$. Moreover, the dipolar
modulation model with an amplitude greater than one does not reproduce
the hemispherical variance asymmetry observed in the CMB. Indeed, the
modulated field has a ring of zero variance instead of the dipolar
asymmetry in this case. The other restriction refers to the index
$\alpha$. In this case, we only consider models for which the dipolar
modulation decreases with the multipole, that is, $\alpha$ is
restricted to be non-negative. Otherwise, the amplitude would be
greater than one for large enough multipoles, resulting again in a
singularity in the covariance matrix and the wrong physical behaviour
as in the previous case. The prior distribution is considered to be
uniform within the intervals described above. These constraints on the
parameters of the model define a improper prior (it is not possible to
normalize the prior to unity because there is no an upper bound in
$\alpha$). This is not a problem for the sampling because we only need
the unnormalized posterior in this case. However, improper priors can
be an issue when different models are compared through the Bayes
factor. In this case, we need to get samples of a (proper) prior in
order to calculate the evidence (see below the discussion on the Bayes
factor).

Since the dipolar modulation is given by a three-dimensional vector,
directional statistics must be applied to the sampled
distributions. This statistics is preferred to study the dipolar
modulation because the significance of the detection can be derived
directly from the deviation of the zero vector. The mean amplitude
vector is defined as the average of the samples of $\mathbf{A}$, that
is,
\begin{equation}
\bar{\mathbf{A}} = \frac{1}{N} \sum_{i=1}^N \mathbf{A}_i \ ,
\end{equation}
where the samples $\mathbf{A}_i$, with $i = 1, \dots, N$, are obtained
from the Markov chains. Notice that this average is directly
calculated from the vectors, not the amplitudes, in order to take into
account the directional character of the dipolar modulation. The mean
amplitude vector is $\bar{\mathbf{A}}=0$ in the case of the isotropic
standard model. However, this is not the case of the mean of the
modulus (not averaged as vectors), which is different from zero even
for the standard model. For this reason, the vector statistics is used
to quantify the dipolar modulation anomaly. Following this idea, the
variance of $\mathbf{A}$ is defined to be:
\begin{equation}
\sigma_\mathbf{A}^2 = \frac{1}{N} \sum_{i=1}^N \left( \mathbf{A}_i -
\bar{\mathbf{A}} \right)^2 \ ,
\end{equation}
where the square means the dot product of vectors.

The directional statistics of the dipolar modulation vector at the
pivot multipole $\ell_0=5$ are given in
table~\ref{tab:vector_stat}. In this case, the distribution is
marginalized over the index $\alpha$. By comparing the modulus of the
vector and its standard deviation, it is obtained that the dipolar
modulation is detected with a significance of $2.5\sigma$.

\begin{table}
\begin{center}
\begin{tabular}{|lcccc|}
\hline
\multicolumn{5}{|c|}{Vector statistics}\\
\hline
& $\bar{A}$ & $\sigma_\mathbf{A}$ & $l$ & $b$ \\
$\mathbf{A}$ & 0.209 & 0.083 & $229^\circ$ & $-40^\circ$ \\
\hline
\end{tabular}
\end{center}
\caption{Statistics of the dipolar modulation vector at $\ell=5$
  marginalizing over the index $\alpha$. The modulus of the mean
  vector ($\bar{A}$) and its direction in terms of the two Galactic
  coordinates ($l$ and $b$) are given. The standard deviation
  $\sigma_A$ is the one obtained from the vector statistics.}
\label{tab:vector_stat}
\end{table}

\begin{figure}
\begin{center}
\includegraphics[width=0.8\textwidth]{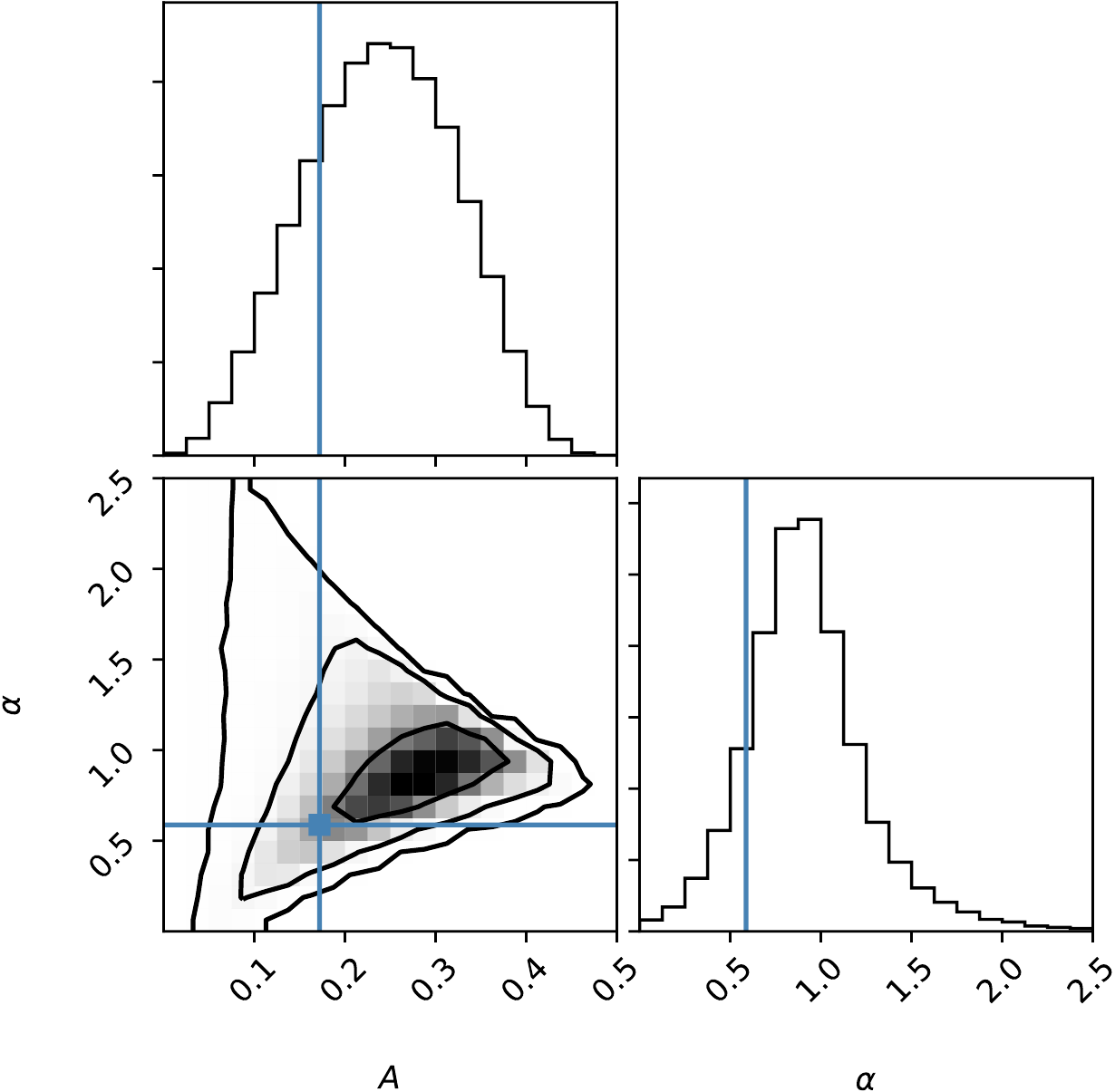}
\includegraphics[width=0.8\textwidth]{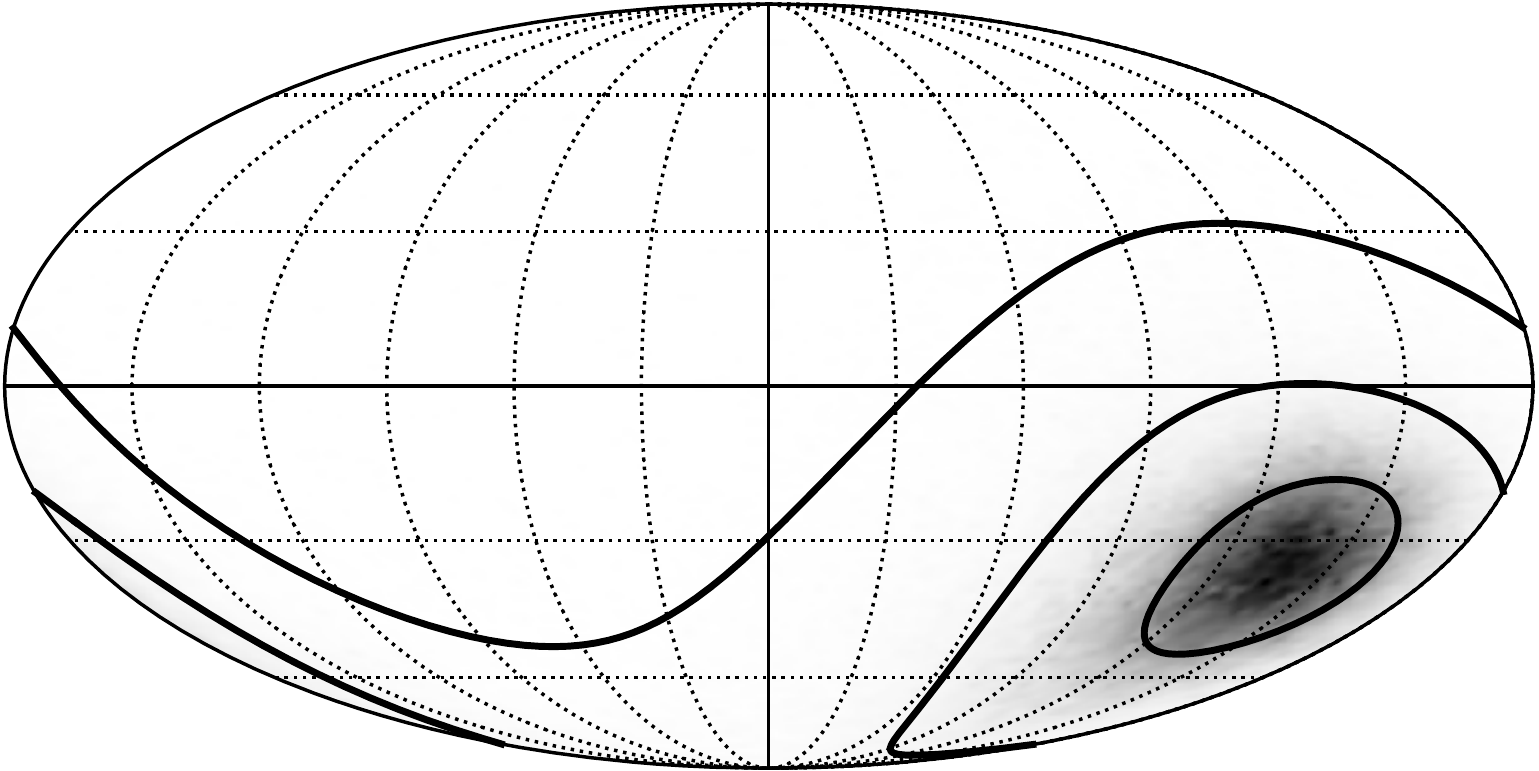}
\end{center}
\caption{\emph{Upper figure:} marginalized posterior distributions for
  the dipolar modulation amplitude at $\ell = 5$ (the pivot multipole)
  and the index $\alpha$ describing the scale dependence. The contours
  in the two-dimensional plot correspond to the 1, 2 and 3$\sigma$
  levels. The blue lines and the dot represent the best-fit parameters
  obtained from the iterative sampling. \emph{Bottom figure:}
  Mollweide projection showing the posterior distribution of the
  dipolar modulation direction. The contours represent the 1, 2 and
  3$\sigma$ confidence levels. Notice that the probabilities
  associated to the contours in the two-dimensional distributions are
  $39\%$, $86\%$ and $99\%$, respectively. The statistics of the
  parameters shown in this figure are given in
  table~\ref{tab:dm_parameters}.}
\label{fig:dm_posterior}
\end{figure}

In figure~\ref{fig:dm_posterior}, the marginalized distributions of
the dipolar modulation parameters are shown. As in the case of the
scale-invariant dipolar modulation, the amplitude is clearly different
from zero indicating the existence of an anisotropic pattern in the
sky. On the other hand, the distribution of the index $\alpha$ shows
that a dipolar modulation with a scale dependence is more likely than
the scale-invariant case. In particular, the value of $\alpha$ is
larger than $0.13$ with a probability of $99.5\%$. The different
statistics of the model parameters shown in
figure~\ref{fig:dm_posterior} are given in
table~\ref{tab:dm_parameters}.

\begin{table}
\begin{center}
\begin{tabular}{|lcccccccccc|}
\hline
\multicolumn{11}{|c|}{Posterior statistics} \\
\hline
         & Mean & Std. & \multicolumn{7}{c}{Percentiles} & Best fit \\ \cline{4-10}
         & & & 0.5\% & 2.5\% & 16\% & 50\% & 84\% &
97.5\% & 99.5\% & \\ \hline
A        & 0.240 & 0.082 & 0.049 & 0.081 & 0.153 & 0.241 & 0.326 &
0.393 & 0.425 & 0.172 \\
$\alpha$ & 0.93 & 0.35 & 0.13 & 0.31 & 0.62 & 0.89 & 1.21 &
1.77 & 2.33 & 0.59 \\
\hline
& \multicolumn{2}{c}{Mean} & \multicolumn{6}{c}{Contours} &
\multicolumn{2}{c|}{Best fit}
\\ \cline{4-9}
& $l$ & $b$ & \multicolumn{2}{c}{39\%} & \multicolumn{2}{c}{86\%} &
\multicolumn{2}{c}{99\%} & $l$ & $b$ \\ \hline
$\mathbf{n}_\mathbf{A}$ & $229^\circ$ & $-40^\circ$ & \multicolumn{2}{c}{$20^\circ$} & \multicolumn{2}{c}{$45^\circ$} &
\multicolumn{2}{c}{$81^\circ$} & $226^\circ$ & $-37^\circ$ \\
\hline
\end{tabular}
\end{center}
\caption{Statistics of the scale-dependent dipolar modulation
  parameters. The corresponding probability distributions are show in
  figure~\ref{fig:dm_posterior}. Whereas the statistics given for the
  amplitude $A$ and the index $\alpha$ are the ones representing the
  one-dimensional marginalized distributions, the dipolar modulation
  direction $\mathbf{n}_\mathbf{A}$ is characterized by the mean
  vector and the two-dimensional contour probabilities. The best-fit
  parameters of the joint posterior distribution are also given in the
  table.}
\label{tab:dm_parameters}
\end{table}

\subsection{Bayesian inference analysis}

Finally, the scale-dependent dipolar modulation model is compared with
the scale-invariant case ($\alpha=0$) previously assumed in the
literature \cite{gordon2007,hoftuft2009}. Additionally, the two
dipolar modulation models are contrasted with the standard isotropic
model. This comparison is done in terms of the Bayesian evidence,
which is given by the mean of the likelihood function over the prior
distribution:
\begin{equation}
Z(\mathbf{d},\hat{\mathbf{d}}) = \int
L(\mathbf{d},\hat{\mathbf{d}}|\theta) \ \pi(\theta) \ \mathrm{d}\theta
\ ,
\label{eqn:evidence}
\end{equation}
where $\hat{\mathbf{d}}$ represents the inpainted data. In order to
calculate the evidence we generate several full-sky simulations
following the probability $p(\hat{\mathbf{d}}|\mathbf{d},\theta_0)$,
where $\theta_0$ are the parameters of some fiducial model (no dipolar
modulation in our case). The likelihood can be easily computed for
these full-sky realizations and the above integral can be calculated
by Monte Carlo integration (the likelihood function is averaged over
samples of the prior distribution). The marginalized evidence
$Z(\mathbf{d})$ can be calculated by the importance sampling method:
\begin{equation}
Z(\mathbf{d}) = \int Z(\mathbf{d},\hat{\mathbf{d}})
\ \mathrm{d}\hat{d} \approx \frac{1}{N} \sum_{i=1}^N
\frac{Z(\mathbf{d},\hat{\mathbf{d}}_i)}{p(\hat{\mathbf{d}}_i|\mathbf{d},\theta_0)}
\ ,
\end{equation}
where the sum is over different inpainted realizations. Notice that
the evidence $Z(\mathbf{d})$ does not depend on the fiducial model
used for generating the inpainted realizations. The information of
$\hat{\mathbf{d}}$ is average out in the above integration.

Since we have to generate prior samples for estimating the evidence,
the prior distribution must be normalized to unity. The normalization
is also important because the evidence given in
eq.~\eqref{eqn:evidence} depends on this normalization factor. The
improper prior probability density considered previously can be
normalized just by considering a maximum allowed index
$\alpha_\mathrm{max}$. In principle, there is not a concrete value for
the maximum index which can be said to be natural, although large
values of $\alpha$ give extremely steep models. We assumed that
$\alpha_\mathrm{max} = 4$ in the calculation of the evidence. Since
the likelihood vanishes for values greater than this index, the
evidence for larger values of $\alpha_\mathrm{max}$ can be estimated
by considering the ratio of indices.\footnote{Assuming that the
  likelihood $L(\alpha)$ vanishes for $\alpha>\alpha_\mathrm{max}$,
  the evidence $Z(\alpha_\mathrm{max})$ calculated with a uniform
  prior in the range $0<\alpha<\alpha_\mathrm{max}$ transforms as
\begin{equation}
Z(\alpha_\mathrm{max}^\prime) =
\frac{\alpha_\mathrm{max}}{\alpha_\mathrm{max}^\prime}
Z(\alpha_\mathrm{max}) \ ,
\end{equation}
with $\alpha_\mathrm{max}^\prime > \alpha_\mathrm{max}$.
}

The Bayes factor $K$ is defined as the ratio of evidences of two
different models. If $K > 1$, the model in the numerator is favoured
with respect to the one in the denominator. Depending on the value of
$K$ the preference of a model against the other is more or less
significant. According to the standard criteria, the Bayes factor
starts to be significant when $\log_{10} K > 0.5$ and it is clearly
decisive when $\log_{10} K > 2$ \cite{jeffreys1961}.

The Bayes factors of the scale-dependent and scale-invariant dipolar
modulation models with respect to the standard isotropic model are
shown in table~\ref{tab:bayes_factor}. The dipolar modulation with
$\alpha=0$ is strongly disfavoured with respect to the standard model
in terms of the Bayesian evidence. This is a reflection of the fact
that the standard model offers a good overall fit to the CMB
temperature data. In principle, it is difficult to find an extension
of the standard model with greater evidence, since the observed
large-scale deviations are dominated by the cosmic variance and their
significance is not high enough to have a large weight in the Bayesian
inference. However, the scale-dependent dipolar modulation has an
evidence comparable to the one obtained from the standard model. This
indicates that the new model of dipolar modulation with scale
dependence agrees with the large-scale CMB temperature data ($\ell \leq
64$) as well as the standard model.

A similar study in terms of the Bayesian evidence of the
scale-invariant dipolar modulation has also been done previously in
\cite{hoftuft2009}. However, in that paper, it is considered a much
smaller prior for the dipolar modulation amplitude ($A<0.15$). If we
rescale the evidences obtained in that paper in order to have the same
prior than the one considered in this work, it is obtained that
$\log_{10} K = -1.34$, which is a strong evidence against the
scale-invariant dipolar modulation with respect to the standard
model. Although this value is greater than the Bayes factor given in
table~\ref{tab:bayes_factor} for the same case ($\log_{10} K =
-2.18$), the two numbers agree on rejecting the scale-invariant
model. The remaining difference between these two Bayes factors could
be caused by the different dipolar modulation model and resolution
considered in \cite{hoftuft2009}.\footnote{Two additional parameters
  are considered in the dipolar modulation model in
  \cite{hoftuft2009}. These parameters modified the isotropic part of
  the model allowing for additional degrees of freedom in the
  variation of the angular power spectrum with multipole. In
  principle, these nuisance parameters may accommodate the observed
  low variance in the largest scales of the CMB.}  Notice that the
dependence of the prior on the maximum amplitude is strong because it
scales with the volume, and therefore, a factor of three must be
included in the volume term. We argue that the prior of the amplitude
assumed in our analysis ($A<1$) only relies on the validity of the
model (the covariance matrix of the modulated field becomes singular
for larger values of the amplitude), and not in a value given ad hoc
or specified in terms of the posterior distribution.

As discussed above, the evidence of the scale-dependent model depends
on the value of $\alpha_\mathrm{max}$ assumed in the prior
distribution. We can always have an arbitrary small evidence by
considering a large enough value of $\alpha_\mathrm{max}$. Therefore,
a maximum value of the index must be taken as the limit of the power
law model. In order to have a Bayes factor supporting the standard
model ($\log_{10} K > 0.5$), the value of the maximum index should be
greater than $\alpha_\mathrm{max} = 12$. In this extreme scenario, the
power law model for the dipolar modulation amplitude gives a
contribution in the multipoles $\ell>2$ less than $1\%$ of the
corresponding amplitude of the quadrupole. Therefore, this class of
models will be better described by a term only modulating the
quadrupole, not a power law model as considered in this work. Since
$\alpha<2.33$ with probability $99.5\%$, models with such a large
value of the index are not favoured by the data (see
table~\ref{tab:dm_parameters}). In the light of this analysis, we
claim that the conclusion that the scale-dependent dipolar modulation
has similar evidence than the standard model is quite robust in spite
of the ambiguity of the evidence due to the dependence on
$\alpha_\mathrm{max}$. Notice that the criteria for model selection in
Bayesian inference are given in terms of $\log_{10} K$, and therefore,
the dependence of this quantity on $\alpha_\mathrm{max}$ is also
logarithmic, giving a weak dependence of the results on the maximum
index considered.

On the other hand, it is found that $\log_{10} K = 2.13$ when
comparing the scale-dependent and the scale-independent scenarios (see
table~\ref{tab:bayes_factor}). This large Bayes factor strongly
supports the dipolar modulation model with scale dependence ($\alpha
\neq 0$) against the standard dipolar modulation with scale-invariant
amplitude.

\begin{table}
\begin{center}
\begin{tabular}{|lccc|}
\hline
 & \multicolumn{2}{c}{Bayes factor with respect to the SM} &
Difference \\ \cline{2-3}
 & scale-dependent DM & scale-invariant DM & \\
\hline
$\log_{10} K$ & -0.057 & -2.18 & 2.13 \\
\hline
\end{tabular}
\end{center}
\caption{Bayes factors of the scale-dependent and scale-invariant
  dipolar modulation models with respect to the standard isotropic
  model. The difference between $\log_{10} K$ of the two model is also
  shown. This difference represents the logarithm of the Bayes factor
  of the scale-dependent model with respect to the scale-invariant one
  with $\alpha=0$. The prior of the index $\alpha$ considered in the
  scale-dependent model is uniform in the range $0<\alpha<4$, that is,
  $\alpha_\mathrm{max} = 4$. As in all the previous analyses, the
  maximum multipole considered in the likelihood is $\ell_\mathrm{max}
  = 64$.}
\label{tab:bayes_factor}
\end{table}

\subsection{CMB temperature without dipolar modulation and the anomaly
  map}

\begin{figure}
\begin{center}
\includegraphics[width=0.49\textwidth]{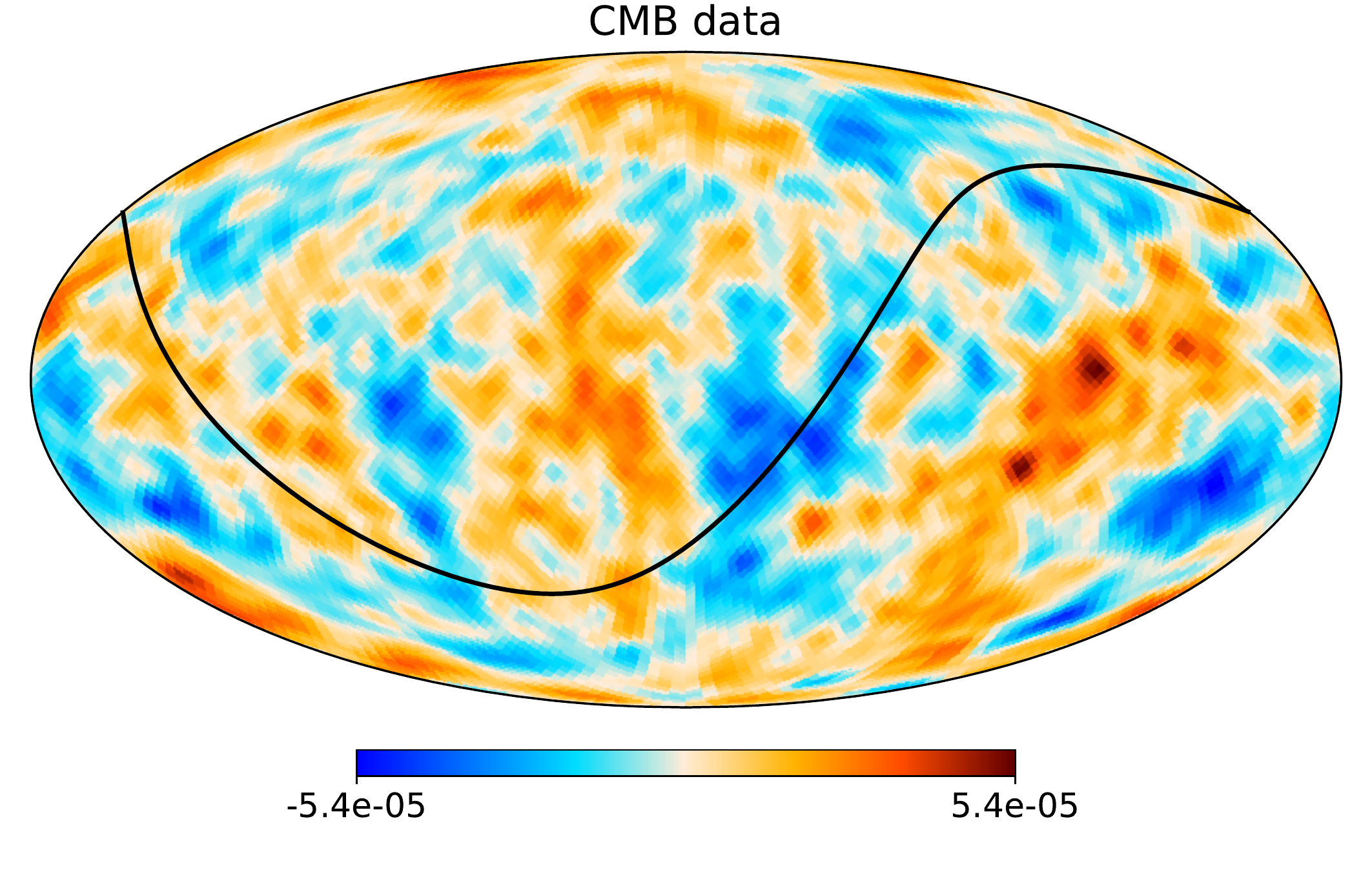}
\end{center}
\begin{center}
\includegraphics[width=0.49\textwidth]{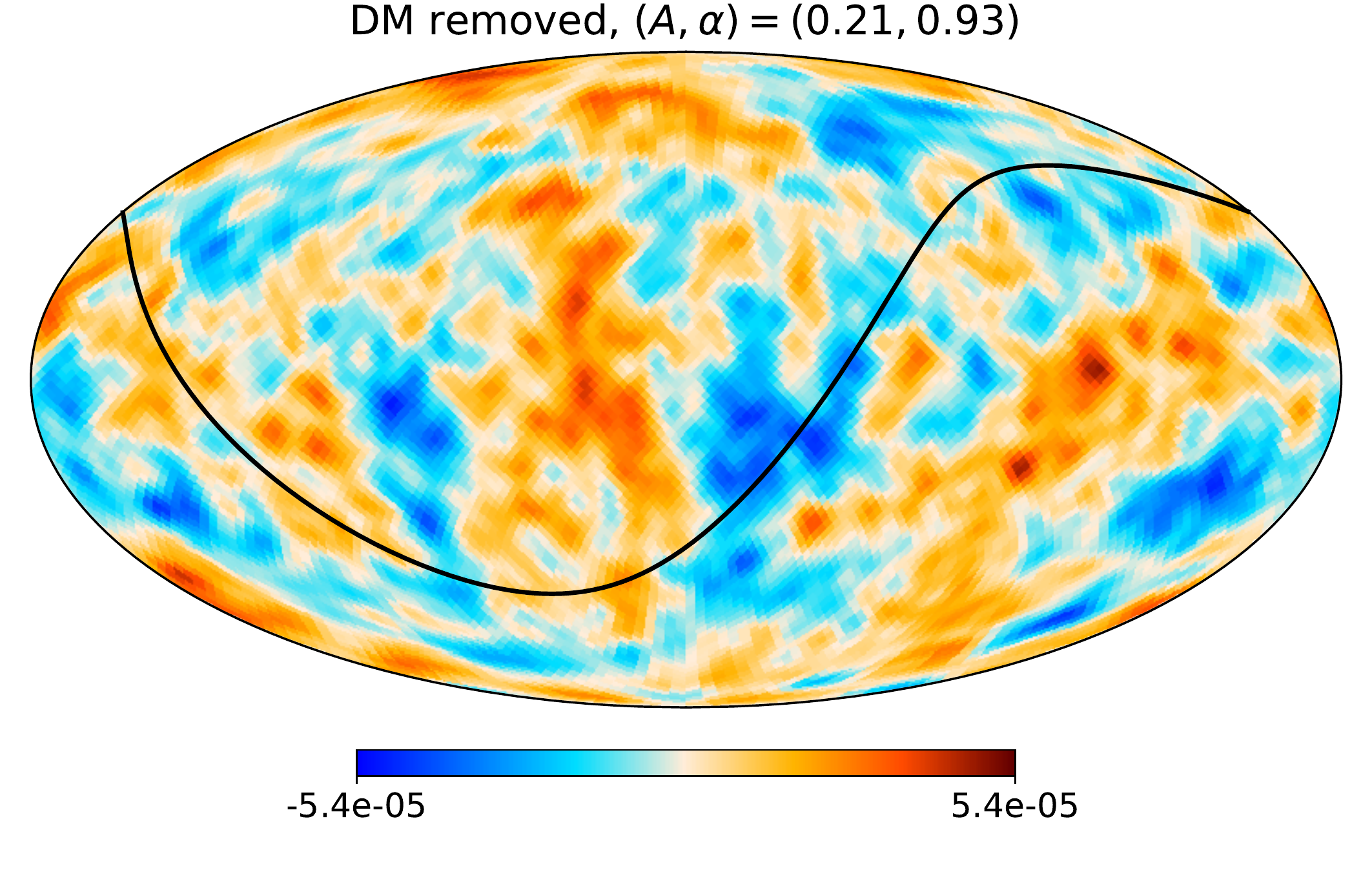}
\includegraphics[width=0.49\textwidth]{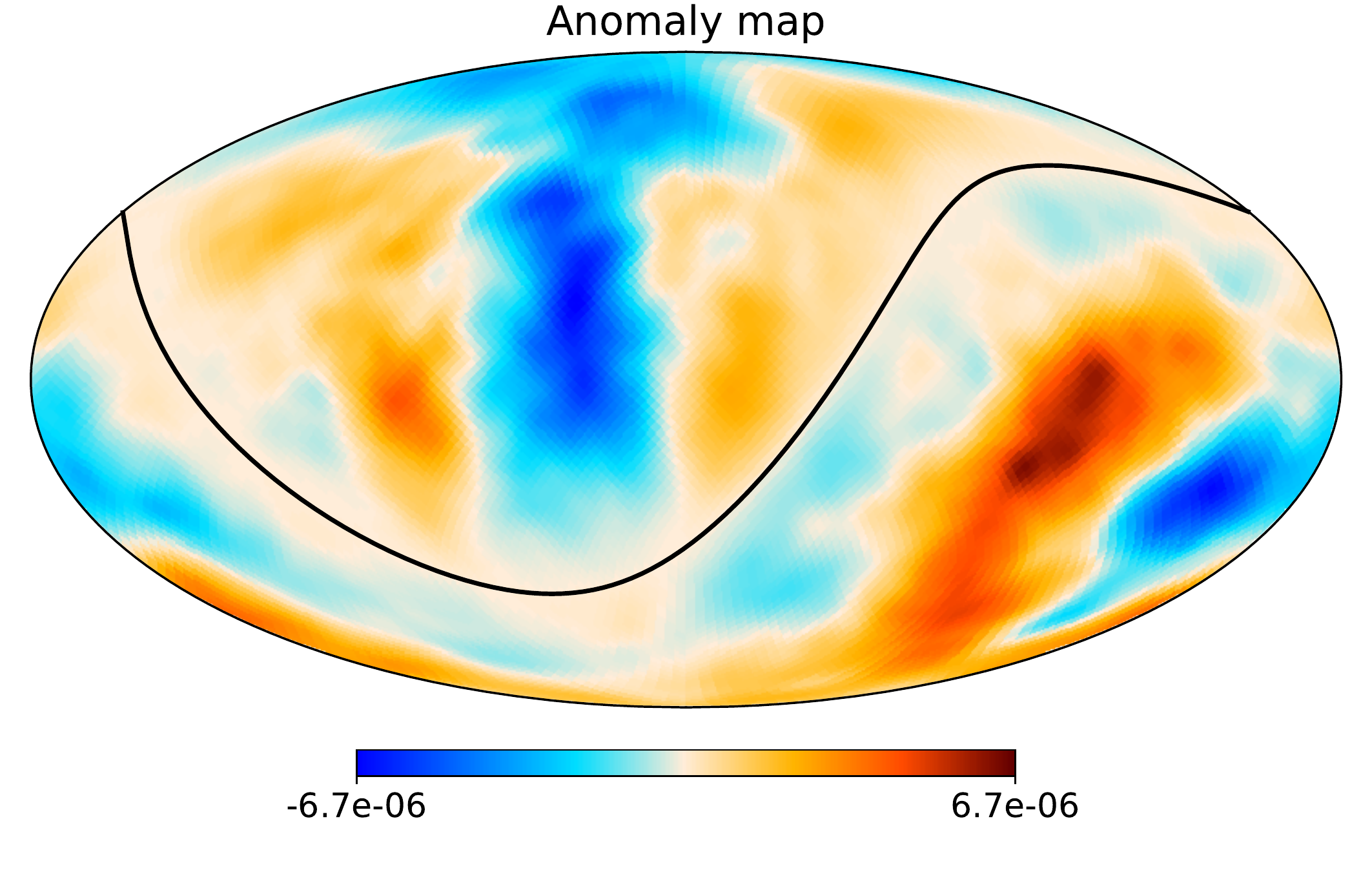}
\end{center}
\caption{\emph{Upper map:} the inpainted CMB temperature data filtered
  with a FWHM of $5^\circ$. \emph{Bottom left map:} the
  scale-dependent dipolar modulation observed in the CMB temperature
  is removed from the data. We assume the dipolar modulation
  parameters given by the vector statistics in
  table~\ref{tab:vector_stat}: $(A, \alpha) = (0.21,
  0.93)$. \emph{Bottom right map:} the dipolar modulation anomaly
  defined as the difference between the observed data and the map
  without the dipolar modulation. The equator of the observed dipolar
  modulation is represented by the black lines.}
\label{fig:cmb_dm_removed}
\end{figure}

Finally, it is possible to remove statistically the observed dipolar
modulation from the data, obtaining a more isotropic temperature
field. This map would be an attempt to obtain a CMB temperature field
which is more compatible with the isotropic $\Lambda$CMB model than
the observed data. The procedure for calculating this isotropic map
consists in applying the same transformation described in
section~\ref{sec:likelihood} used for calculating the likelihood. The
relation between the spherical harmonics coefficients given in
eq.~\eqref{eqn:hat_alm} has to be inverted in order to calculate the
isotropic coefficients $a_{\ell m}$ from the ones characterizing the
field with dipolar modulation ($\hat{a}_{\ell m}$). (see
appendix~\ref{app:inversion_alm} for the details of the
inversion). The dipolar modulation parameters assumed in the inversion
are the ones given in table~\ref{tab:vector_stat}, that is, $(A,
\alpha) = (0.21, 0.93)$. The resulting maps of this procedure are
shown in figure~\ref{fig:cmb_dm_removed}. We can see that, once the
dipolar modulation is removed, the temperature field seems to be more
isotropic. The anomaly map is defined to be the difference between the
observed CMB data and the isotropic estimation obtained by removing
the dipolar modulation. This map shows clearly a large-scale pattern,
indicating that most of the dipolar modulation anomaly affects the low
multipoles.

\section{Quadrupole and octopole alignment}
\label{sec:qo_alignment}

Several estimators have been proposed to quantify the
quadrupole-octopole alignment. Here, we discuss three of them
previously used in the literature: one based on the angular momentum
dispersion and the other two in the multipole vectors formalism.

\subsection{Maximum angular momentum dispersion}

One of the first estimators proposed in the literature for quantifying
the alignment between the quadrupole and the octopole is based on
planes with maximum variance \cite{deOliveira-Costa2004}. Considering
the unit vector $\mathbf{n}$, the angular momentum is projected along
the direction given by this vector ($\mathbf{n} \cdot
\mathbf{L}$). The variance of this quantity measures the amplitude of
the fluctuations on planes perpendicular to $\mathbf{n}$.\footnote{In
  the particular case when the $z$-direction is given by $\mathbf{n}$,
  it is possible to see that the operator $\mathbf{n} \cdot
  \mathbf{L}$ is proportional to the derivative with respect to the
  azimuthal angle $\phi$. Therefore, the variance of this quantity
  measures how much the field varies on the direction perpendicular to
  $\mathbf{n}$.}  Decomposing the temperature field in the different
multipolar moments $T_\ell$, the variance of the angular momentum
projected on the $\mathbf{n}$-direction as a function of $\ell$ is
given by
\begin{equation}
\sum_{m=-\ell}^\ell m^2 |a_{\ell m}(\mathbf{n})|^2 \ ,
\label{eqn:maxL_estimator}
\end{equation}
where $a_{\ell m}(\mathbf{n})$ are the spherical harmonic coefficients
calculated in the system of reference in which the $z$-axis is
parallel to $\mathbf{n}$. Given the spherical harmonic coefficients in
an arbitrary system of reference, the coefficients $a_{\ell
  m}(\mathbf{n})$ can be calculated by applying the appropriate Wigner
D-matrix. Finally, the direction $\mathbf{n}_\ell$ characterizing the
multipole $\ell$ is calculated by maximizing the angular momentum
variance given in eq.~\eqref{eqn:maxL_estimator}.

Since different multipoles are independent in an isotropic field, the
directions given by $\mathbf{n}_\ell$ would also be independent. If
this is the case, the cosine of the angle between two directions,
$\mathbf{n}_\ell \cdot \mathbf{n}_{\ell^\prime}$, is uniformly
distributed in the interval $[-1,1]$. Notice that the quantity given
in eq.~\eqref{eqn:maxL_estimator} is invariant under parity
transformation $\mathbf{n} \to -\mathbf{n}$, and therefore, the sign
of $\mathbf{n}_\ell \cdot \mathbf{n}_{\ell^\prime}$ is not
relevant. In the present analysis, we quantify the quadrupole-octopole
alignment with the estimator $|\mathbf{n}_2 \cdot \mathbf{n}_3|$,
which is uniformly distributed in the range $[0,1]$ according to the
standard model. Values of $|\mathbf{n}_2 \cdot \mathbf{n}_3|$ very
close to $1$ may imply an unexpected alignment between these two
multipoles.

\subsection{Multipole vectors}

In addition to the standard expansion in terms of the spherical
harmonics, there are other ways to represent a field on the
sphere. One of them used in isotropy analysis of the CMB is the
multipole vectors decomposition \cite{weeks2004,copi2004}. In this
formalism, the $\ell$-component of the temperature field $T_\ell$ is
expressed as a function of $\ell$ unit vectors, $\mathbf{v}^1_\ell
\dots \mathbf{v}^\ell_\ell$, and an amplitude $A_\ell$ \ .
\begin{equation}
T_\ell(\mathbf{n}) = A_\ell \left[ \left( \mathbf{v}^1_\ell \cdot
  \mathbf{n} \right) \dots \left( \mathbf{v}^\ell_\ell \cdot
  \mathbf{n} \right) - B_\ell(\mathbf{n}) \right] \ .
\end{equation}
where $B_\ell$ removes the contribution of lower order multipoles from
the preceding product. Notice that, in the case of the dipole, all the
information is given by a vector, $\mathbf{A}_1 \equiv A_1
\mathbf{v}_1^1$, as it is expected from the standard decomposition in
terms of the spherical harmonics.\footnote{In the case of the dipole,
  the components of the vector $\mathbf{A}_1$ are given by the
  following expressions in terms of the spherical harmonics
  coefficients:
\begin{subequations}
\begin{equation}
A_x = \sqrt{\frac{3}{4\pi}} \frac{1}{\sqrt{2}} \left( a_{1-1} -
a_{11}\right) \ ,
\end{equation}
\begin{equation}
A_y = \sqrt{\frac{3}{4\pi}} \frac{1}{i\sqrt{2}} \left( a_{1-1} +
a_{11}\right) \ ,
\end{equation}
\begin{equation}
A_z = \sqrt{\frac{3}{4\pi}} a_{10} \ .
\end{equation}
\end{subequations}
} The multipole vectors approach generalizes this well-known
characterization of the dipole to higher-order multipoles.

In this scheme, the isotropy of the field can be tested by estimating
the amount of alignment between different multipole vectors. Since
there are many vectors for each multipole (two representing the
quadrupole, and three for the octopole), several combinations of
vectors can be considered. In this work, we use two alignment
estimators that have been used previously in the literature
\cite{copi2015}.

Given two multipole vectors, the area vector is defined by the cross
product:
\begin{equation}
\mathbf{w}_\ell^{ij} = \mathbf{v}_\ell^i \times \mathbf{v}_\ell^j \ .
\end{equation}
In our case, we are only interested in the quadrupole and the
octopole, and hence, there is only an area vector for the quadrupole
and three vectors for the octopole:
\begin{subequations}
\begin{equation}
\mathbf{w}_2 \equiv \mathbf{w}_2^{12} \ ,
\end{equation}
\begin{equation}
\mathbf{w}_3^1 \equiv \mathbf{w}_3^{23} \ , \quad \mathbf{w}_3^2
\equiv \mathbf{w}_3^{31} \ , \quad \mathbf{w}_3^3 \equiv
\mathbf{w}_3^{12} \ .
\end{equation}
\end{subequations}
Notice that the vector $\mathbf{w}_2$ derived from the quadrupole is
the same than the one obtained with the maximum angular momentum
dispersion ($\mathbf{n}_2$ in the previous section). However, there is
no a direct correspondence between the vectors $\mathbf{w}_3^i$ and
$\mathbf{n}_3$ in the case of the octopole \cite{copi2006}.

In general, the alignment of these vectors with an arbitrary direction
can be quantified by the dot product. Therefore, the
quadrupole-octopole alignment is characterized by the three following
quantities:
\begin{equation}
A_i = |\mathbf{w}_2 \cdot \mathbf{w}_3^i| \ .
\end{equation}
In order to combine all the information, two estimators are proposed
\cite{copi2015}:
\begin{subequations}
\begin{equation}
S \equiv \frac{1}{3} \sum_{i=1}^3 A_i \ ,
\end{equation}
\begin{equation}
T \equiv 1 - \frac{1}{3}
\sum_{i=1}^3 \left( 1 - A_i \right)^2 \ .
\end{equation}
\end{subequations}
Whereas the $S$ estimator represents the average of the alignment
vectors, $T$ is a quadratic combination of these vectors. Both
estimators take values in the range $[0,1]$, being $1$ the maximum
alignment.

\section{Results on the quadrupole-octopole alignment}
\label{sec:results_qo}

In this work, the implications of the scale-dependent dipolar
modulation model on the alignment of the quadrupole and octopole are
explored. From eq.~\eqref{eqn:hat_alm}, it can be shown that the
dipolar modulation adds correlations between consecutive multipoles,
then a quadrupole-octopole correlation is expected. If the amplitude
of the dipolar modulation is large enough for the smallest multipoles,
then alignment between the quadrupole and the octopole may be
explained within this sort of non-isotropic models. In this section,
we quantify the impact of the scale-dependent dipolar modulation model
proposed in this work on the quadrupole-octopole alignment.

Once the parameters of the dipolar modulation model are sampled
according to the posterior, these samples are used as random
realizations of a non-isotropic Universe. A CMB temperature field with
a dipolar modulation is generated for each set of parameters following
eq.~\eqref{eqn:dm_model}. Hence, we obtain a sample of CMB simulations
for the scale-dependent dipolar modulation model compatible with the
posterior. Finally, the probability distribution of the
quadrupole-octopole alignment estimators are calculated from theses
samples.

\begin{figure}
\begin{center}
\includegraphics[width=0.5\textwidth]{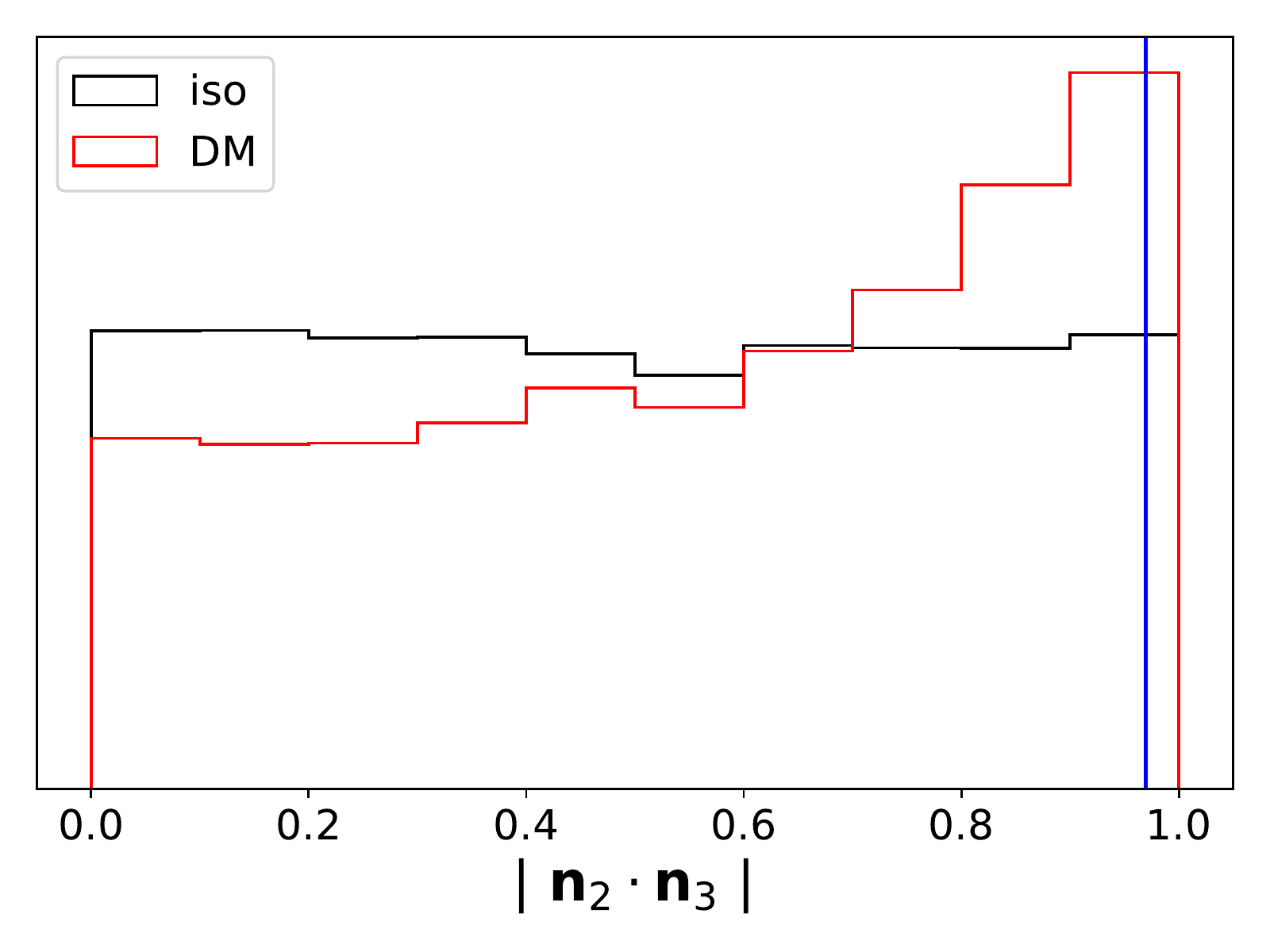}
\break
\includegraphics[width=0.49\textwidth]{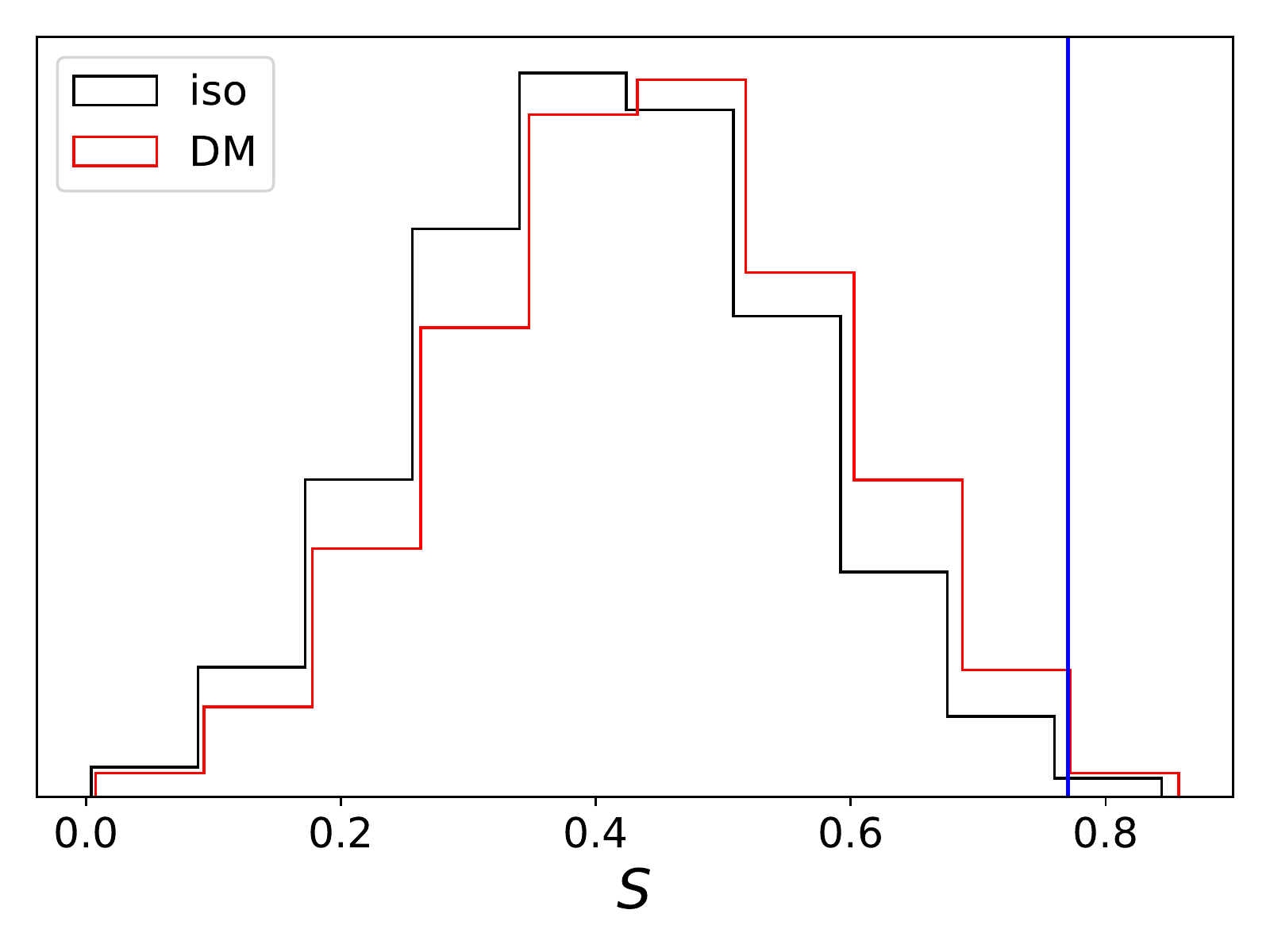}
\includegraphics[width=0.49\textwidth]{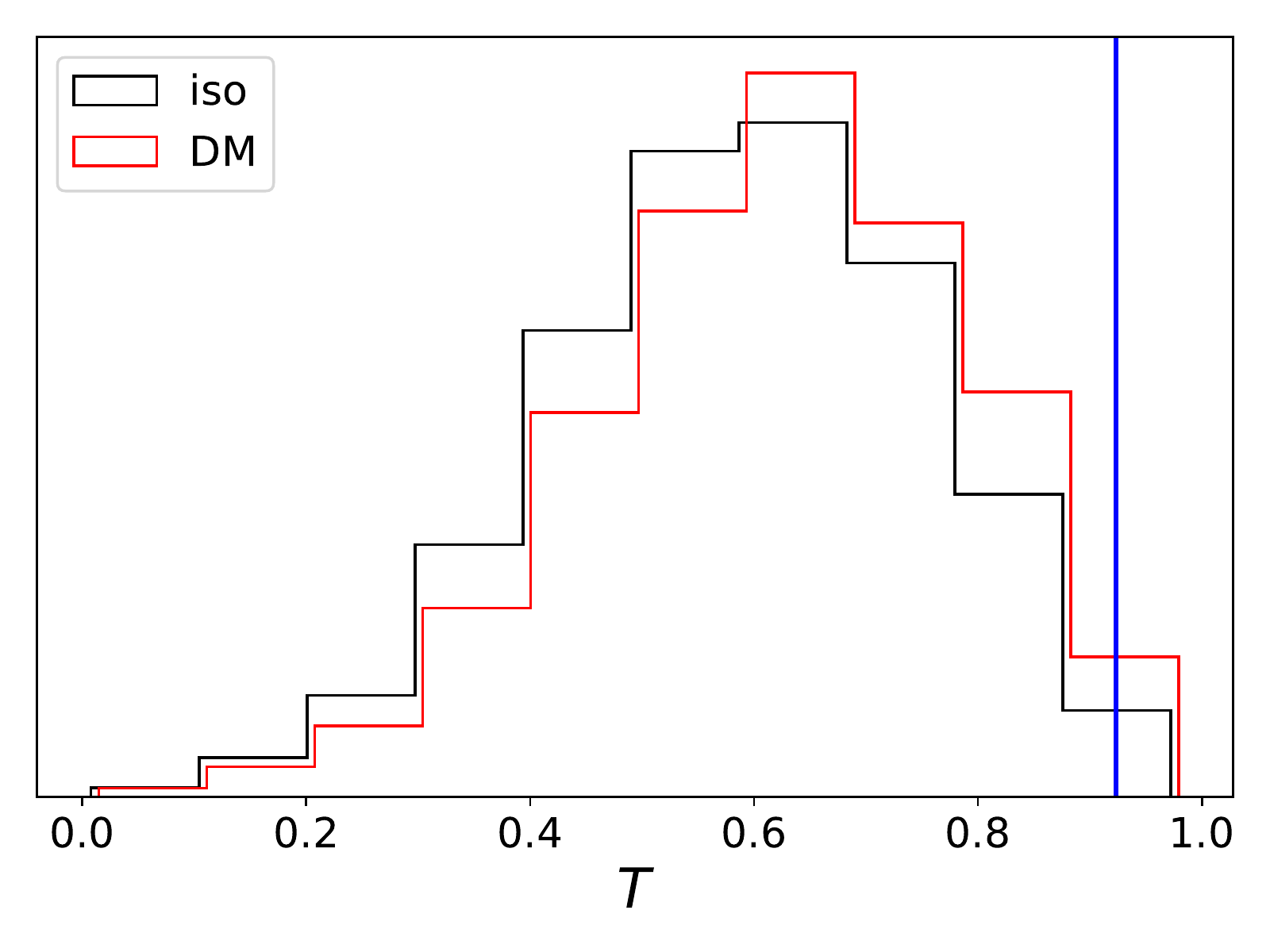}
\end{center}
\caption{Probability density distributions of the three
  quadrupole-octopole alignment estimators considered in this
  work. Whereas the black histograms represent the distribution
  obtained assuming the standard isotropic model, the red curves
  correspond to the probabilities in the case of having a
  scale-dependent dipolar modulation. The values estimated from the
  data are shown by the blue vertical lines.}
\label{fig:qo_dist}
\end{figure}

In figure~\ref{fig:qo_dist}, the probability distributions of the
three alignment estimators analyzed in this work are shown. It is
possible to see that the alignment between the quadrupole and the
octopole is favoured under the hypothesis that there exists a
scale-dependent dipolar modulation with respect to the standard
isotropic case. This is particularly evident for $|\mathbf{n}_2 \cdot
\mathbf{n}_3|$, where a clear deviation from the uniform distribution
is shown. In the three estimators, values close to one are more likely
when the CMB temperature has a scale-dependent dipolar modulation. The
$p$-values obtained in each case are shown in
table~\ref{tab:qo_pvalue}. It is found that the alignment probability
is larger for the dipolar modulation model than for the isotropic case
by a factor of $\approx 1.8$. Only the estimator $S$ remains below the
$1\%$ level, although its $p$-value has also increased by a similar
factor than the others. Notice that this statistical dependence of the
two anomalies is only observed if the dipolar modulation is considered
to be scale-dependent. This correlation is not observed in the
scale-invariant case \cite{polastri2015}.

\begin{table}
\begin{center}
\begin{tabular}{|lcc|}
\hline
 & \multicolumn{2}{c|}{$p$-value} \\
 & Isotropic & scale-dependent DM \\
\hline
$\mathbf{n}_2 \cdot \mathbf{n}_3$ & $0.0287$ & $0.0540$ \\
$S$                               & $0.0045$ & $0.0079$ \\
$T$                               & $0.0076$ & $0.0140$ \\
\hline
\end{tabular}
\end{center}
\caption{The $p$-values for the three quadrupole-octopole alignment
  estimators considered in this work. The corresponding values
  assuming both the isotropic and the scale-dependent dipolar
  modulation models are shown.}
\label{tab:qo_pvalue}
\end{table}

\section{Conclusions}
\label{sec:conclusions}

One of the open questions on CMB physics are the nature of the
different large-scale deviations from the standard model observed in
the data. In this work, we study the influence of the dipolar
modulation on the quadrupole-octopole alignment. Previous analysis
have shown that there is a negligible correlation between both
anomalies \cite{polastri2015}. However, the dipolar modulation model
considered in those works assumes that the amplitude of the modulation
is independent on the scale. Signatures of a scale-dependent
modulation have been reported in \cite{planck162015} by studying the
dipolar modulation in different multipole intervals. In this paper, we
propose a scale-dependent dipolar modulation model which is more
consistent with the data and reproduces the observed large-scale
temperature fluctuations. We find that the Bayesian evidence supports
the scale-dependent model versus the one with scale-invariant
amplitude. Additionally, if the scale-dependent dipolar modulation
model is compared to the standard model, similar evidences are
obtained (Bayes factor $K \approx 1$). Therefore, this new model of
dipolar modulation represents a good fit to the CMB data at large
scales ($\ell \leq 64$), at least, as well as the standard model.

Once a suitable dipolar modulation model is adopted, the alignment
between the quadrupole and the octopole is analyzed. Three different
estimators commonly used in the literature for quantifying the
alignment are studied. The corresponding statistical distributions are
calculated under the hypothesis that a scale-dependent dipolar
modulation may exists. For this purpose, the parameters of the dipolar
model are sampled according to the posterior distribution. Since the
dipolar model favours that some region of the sky has more large-scale
power than the opposite one, the low multipoles of the CMB must be
correlated in a particular way. Therefore, the alignment between the
large-scale multipoles is more likely in the dipolar modulation
scenario than in the standard model. However, this can be only achieve
considering the scale-dependent dipolar modulation. This model makes
the distributions of the alignment estimators change with respect to
the isotropic case so that the quadrupole-octopole alignment is more
likely. We find an increment of a factor of $1.80$ ($80 \%$) in the
p-value in all the estimators considered. However, there is still a
deviation whose $p$-value is below $1 \%$ in one of the estimators
($S$) based on multipole vectors.

Notice that, in this work, we have only considered the possibility
that a particular dipolar modulation model {could explain} the
quadrupole-octopole alignment. However, it is possible that a more
general model accounts for several of the large-scale CMB anomalies at
the same time reducing the joint significance. For the moment, we have
shown that a more realistic dipolar modulation including the scale
dependence is clearly correlated with the large-scale CMB alignments,
and therefore, these two anomalies are not independent. Further work
may consists in finding better models for explaining the observed
large-scale deviations in the CMB within a more physical frame.

\appendix

\section{Scale-dependent dipolar modulation in spherical harmonics}
\label{app:dm_spherical_harmonics}

In order to calculate the likelihood function or to reconstruct the
isotropy map from the data, it is useful to work in the spherical
harmonics space. In this appendix, we calculate the spherical
harmonics transform of the dipolar modulation model with
scale-dependent amplitude. Similar calculation was done before for the
constant amplitude model in \cite{moss2011}. For simplicity, we
restrict our calculation to the case in which the dipolar modulation
direction coincides with the $z$-axis. The dipolar model in
eq.~\eqref{eqn:dm_model} in spherical coordinates is given by:
\begin{equation}
\hat{T}(\theta,\phi) = \left[ 1 + \cos \theta
  \ A(\theta) \otimes \right] T(\theta,\phi) \ ,
\label{eqn:dm_model_zaxis}
\end{equation}
where $A(\theta)$ is the filter whose coefficients are the
scale-dependent dipolar amplitudes $A_\ell$ given by
eq.~\eqref{eqn:Al} for the model considered in this work. In order to
calculate the spherical harmonics coefficients of the product of two
functions on the sphere, we use the following identity:
\begin{multline}
\int Y_{\ell_1 m_1} Y_{\ell_2 m_2} Y_{\ell_3 m_3}^* \ \mathrm{d}^2n =
\\ =
(-1)^{m_3} \sqrt{\frac{(2\ell_1+1)(2\ell_2+1)(2\ell_3+1)}{4\pi}}
\left( \begin{array}{ccc}
\ell_1 & \ell_2 & \ell_3 \\ 0 & 0 & 0 \\
\end{array} \right)
\left( \begin{array}{ccc}
\ell_1 & \ell_2 & \ell_3 \\ m_1 & m_2 & -m_3 \\
\end{array} \right) \ .
\end{multline}
The spherical harmonics transform of eq.~\eqref{eqn:dm_model_zaxis}
can be calculated from the previous integral and taking into account
that $\cos \theta = \sqrt{\frac{4\pi}{3}} Y_{10}$:
\begin{multline}
\hat{a}_{\ell m} = a_{\ell m} + \\ + (-1)^m \sum_{\ell^\prime=0}^\infty
\sum_{m^\prime=-\ell^\prime}^{\ell^\prime}
\sqrt{(2\ell+1)(2\ell^\prime+1)}
\left( \begin{array}{ccc}
\ell^\prime & 1 & \ell \\ 0 & 0 & 0 \\
\end{array} \right)
\left( \begin{array}{ccc}
\ell^\prime & 1 & \ell \\ m^\prime & 0 & -m \\
\end{array} \right) A_{\ell^\prime} a_{\ell^\prime m^\prime} \ ,
\end{multline}
where it has been used that the convolution in
eq.~\eqref{eqn:dm_model_zaxis} in Fourier space is just given by the
product of the filter $A_\ell$ and the spherical harmonics
coefficients $a_{\ell m}$. This equation can be simplified using the
selection rules of the Wigner 3-$j$ symbols. In this case, only the
terms with $\ell^\prime=\ell+1$ or $\ell^\prime=\ell-1$ contributes to the
sum. In this case, the Wigner 3-$j$ symbols in the expression above
are given by the identity
\begin{equation}
\left( \begin{array}{ccc}
\ell+1 & 1 & \ell \\ m & 0 & -m \\
\end{array} \right) = 
(-1)^{\ell+m+1} \sqrt{\frac{\ell+1}{(2\ell+1)(2\ell+3)}} \sqrt{1 - \left(
  \frac{m}{\ell+1} \right)^2} \ .
\end{equation}
Finally, we obtain the spherical harmonics coefficients $\hat{a}_{\ell
m}$ of the modulated field in terms of the coefficients $a_{\ell m}$
characterizing the isotropic part
\begin{equation}
\hat{a}_{\ell m} = a_{\ell m} + F_{\ell-1 m} A_{\ell-1} a_{\ell-1 m} +
F_{\ell m} A_{\ell+1} a_{\ell+1 m} \ ,
\end{equation}
where the coefficients $F_{\ell m}$ are
\begin{equation}
F_{\ell m} = \frac{\ell+1}{\sqrt{(2\ell+1)(2\ell+3)}}
\sqrt{1-\left(\frac{m}{\ell+1}\right)^2} \ .
\end{equation}
These equations are the same that the ones in eqs.~\eqref{eqn:hat_alm}
and \eqref{eqn:Flm} and they are the generalization of the expressions
in \cite{moss2011} to the case of scale-dependent dipolar modulation.

\section{Recursive estimation of the isotropic spherical harmonics coefficients}
\label{app:inversion_alm}

The inversion of the system of equations in eq.~\eqref{eqn:hat_alm}
can be done by introducing the following variables, which are
calculated by recursion on $\ell$:
\begin{subequations}
\begin{equation}
u_{\ell m} = F_{\ell m} A_{\ell}/v_{\ell-1 m} \ ,
\end{equation}
\begin{equation}
v_{\ell m} = 1 - u_{\ell m} F_{\ell m} A_{\ell+1} \ ,
\end{equation}
\end{subequations}
with $u_{m m} = 0$ and $v_{m m} = 1$. Likewise, the spherical
harmonics $a_{\ell m}$ of the isotropic field can be calculated from
the modulated coefficients $\hat{a}_{\ell m}$ by solving two recursive
equations. First, the auxiliary coefficients $x_{\ell m}$ are computed
by
\begin{equation}
x_{\ell m} = \hat{a}_{\ell m} - u_{\ell m} x_{\ell-1 m} \ ,
\end{equation}
with the initial condition $x_{m m} = \hat{a}_{m m}$. Finally, the
coefficients $a_{\ell m}$ are given by
\begin{equation}
a_{\ell m} = \left( x_{\ell m} - F_{\ell m} A_{\ell+1} a_{\ell+1 m}
\right) / v_{\ell m} \ ,
\end{equation}
which is solved backwards with the initial condition
$a_{\ell_\mathrm{max} m} = x_{\ell_\mathrm{max} m } /
v_{\ell_\mathrm{max} m}$.

Additionally, the Jacobian of the transformations in
eq.~\eqref{eqn:hat_alm} can also be obtained from the variables
defined above:
\begin{equation}
\log J = \sum_{\ell=0}^{\ell_\mathrm{max}} \sum_{m=0}^\ell \log |
v_{\ell m} | \ .
\end{equation}

\acknowledgments

AMC would like to thank Universidad de Cantabria for a post-doctoral
contract. AMC and EMG acknowledge financial support from Agencia
Estatal de Investigaci\'on (AEI) and Fondo Europeo de Desarrollo
Regional (FEDER, UE), projects ref. ESP2017-83921-C2-1-R and
AYA2017-90675-REDC.

\bibliographystyle{JHEP}
\bibliography{dm.bib}

\providecommand{\href}[2]{#2}\begingroup\raggedright\begin{thebibliography}{10}

\bibitem{planck012018}
{Planck Collaboration} and Y.~{Akrami et al.}, \emph{{Planck 2018 results. I.
  Overview and the cosmological legacy of Planck}}, {\emph{arXiv e-prints}
  (2018) arXiv:1807.06205} [\href{https://arxiv.org/abs/1807.06205}{{\ttfamily
  1807.06205}}].

\bibitem{eriksen2004}
H.~K. {Eriksen}, F.~K. {Hansen}, A.~J. {Banday}, K.~M. {G{\'o}rski} and P.~B.
  {Lilje}, \emph{{Asymmetries in the Cosmic Microwave Background Anisotropy
  Field}}, \href{https://doi.org/10.1086/382267}{\emph{\apj} {\bfseries 605}
  (2004) 14} [\href{https://arxiv.org/abs/astro-ph/0307507}{{\ttfamily
  astro-ph/0307507}}].

\bibitem{hansen2009}
F.~K. {Hansen}, A.~J. {Banday}, K.~M. {G{\'o}rski}, H.~K. {Eriksen} and P.~B.
  {Lilje}, \emph{{Power Asymmetry in Cosmic Microwave Background Fluctuations
  from Full Sky to Sub-Degree Scales: Is the Universe Isotropic?}},
  \href{https://doi.org/10.1088/0004-637X/704/2/1448}{\emph{\apj} {\bfseries
  704} (2009) 1448} [\href{https://arxiv.org/abs/0812.3795}{{\ttfamily
  0812.3795}}].

\bibitem{akrami2014}
Y.~{Akrami}, Y.~{Fantaye}, A.~{Shafieloo}, H.~K. {Eriksen}, F.~K. {Hansen},
  A.~J. {Banday} et~al., \emph{{Power Asymmetry in WMAP and Planck Temperature
  Sky Maps as Measured by a Local Variance Estimator}},
  \href{https://doi.org/10.1088/2041-8205/784/2/L42}{\emph{\apj} {\bfseries
  784} (2014) L42} [\href{https://arxiv.org/abs/1402.0870}{{\ttfamily
  1402.0870}}].

\bibitem{planck232013}
{Planck Collaboration} and P.~A.~R. {Ade et al.}, \emph{{Planck 2013 results.
  XXIII. Isotropy and statistics of the CMB}},
  \href{https://doi.org/10.1051/0004-6361/201321534}{\emph{\aap} {\bfseries
  571} (2014) A23} [\href{https://arxiv.org/abs/1303.5083}{{\ttfamily
  1303.5083}}].

\bibitem{planck162015}
{Planck Collaboration} and P.~A.~R. {Ade et al.}, \emph{{Planck 2015 results.
  XVI. Isotropy and statistics of the CMB}},
  \href{https://doi.org/10.1051/0004-6361/201526681}{\emph{\aap} {\bfseries
  594} (2016) A16} [\href{https://arxiv.org/abs/1506.07135}{{\ttfamily
  1506.07135}}].

\bibitem{planck072018}
{Planck Collaboration} and Y.~{Akrami et al.}, \emph{{Planck 2018 results. VII.
  Isotropy and Statistics of the CMB}}, {\emph{arXiv e-prints} (2019)
  arXiv:1906.02552} [\href{https://arxiv.org/abs/1906.02552}{{\ttfamily
  1906.02552}}].

\bibitem{hoftuft2009}
J.~{Hoftuft}, H.~K. {Eriksen}, A.~J. {Banday}, K.~M. {G{\'o}rski}, F.~K.
  {Hansen} and P.~B. {Lilje}, \emph{{Increasing Evidence for Hemispherical
  Power Asymmetry in the Five-Year WMAP Data}},
  \href{https://doi.org/10.1088/0004-637X/699/2/985}{\emph{\apj} {\bfseries
  699} (2009) 985} [\href{https://arxiv.org/abs/0903.1229}{{\ttfamily
  0903.1229}}].

\bibitem{muir2018}
J.~{Muir}, S.~{Adhikari} and D.~{Huterer}, \emph{{Covariance of CMB
  anomalies}}, \href{https://doi.org/10.1103/PhysRevD.98.023521}{\emph{\prd}
  {\bfseries 98} (2018) 023521}
  [\href{https://arxiv.org/abs/1806.02354}{{\ttfamily 1806.02354}}].

\bibitem{deOliveira-Costa2004}
A.~{de Oliveira-Costa}, M.~{Tegmark}, M.~{Zaldarriaga} and A.~{Hamilton},
  \emph{{Significance of the largest scale CMB fluctuations in WMAP}},
  \href{https://doi.org/10.1103/PhysRevD.69.063516}{\emph{\prd} {\bfseries 69}
  (2004) 063516} [\href{https://arxiv.org/abs/astro-ph/0307282}{{\ttfamily
  astro-ph/0307282}}].

\bibitem{copi2015}
C.~J. {Copi}, D.~{Huterer}, D.~J. {Schwarz} and G.~D. {Starkman},
  \emph{{Large-scale alignments from WMAP and Planck}},
  \href{https://doi.org/10.1093/mnras/stv501}{\emph{\mnras} {\bfseries 449}
  (2015) 3458} [\href{https://arxiv.org/abs/1311.4562}{{\ttfamily 1311.4562}}].

\bibitem{copi2006}
C.~J. {Copi}, D.~{Huterer}, D.~J. {Schwarz} and G.~D. {Starkman}, \emph{{On the
  large-angle anomalies of the microwave sky}},
  \href{https://doi.org/10.1111/j.1365-2966.2005.09980.x}{\emph{\mnras}
  {\bfseries 367} (2006) 79}
  [\href{https://arxiv.org/abs/astro-ph/0508047}{{\ttfamily
  astro-ph/0508047}}].

\bibitem{marcos-caballero2017}
A.~{Marcos-Caballero}, E.~{Mart{\'\i}nez-Gonz{\'a}lez} and P.~{Vielva},
  \emph{{Local properties of the large-scale peaks of the CMB temperature}},
  \href{https://doi.org/10.1088/1475-7516/2017/05/023}{\emph{Journal of
  Cosmology and Astro-Particle Physics} {\bfseries 2017} (2017) 023}
  [\href{https://arxiv.org/abs/1701.08552}{{\ttfamily 1701.08552}}].

\bibitem{gordon2007}
C.~{Gordon}, \emph{{Broken Isotropy from a Linear Modulation of the Primordial
  Perturbations}}, \href{https://doi.org/10.1086/510511}{\emph{\apj} {\bfseries
  656} (2007) 636} [\href{https://arxiv.org/abs/astro-ph/0607423}{{\ttfamily
  astro-ph/0607423}}].

\bibitem{moss2011}
A.~{Moss}, D.~{Scott}, J.~P. {Zibin} and R.~{Battye}, \emph{{Tilted physics: A
  cosmologically dipole-modulated sky}},
  \href{https://doi.org/10.1103/PhysRevD.84.023014}{\emph{\prd} {\bfseries 84}
  (2011) 023014} [\href{https://arxiv.org/abs/1011.2990}{{\ttfamily
  1011.2990}}].

\bibitem{planck042018}
{Planck Collaboration} and Y.~{Akrami et al.}, \emph{{Planck 2018 results. IV.
  Diffuse component separation}}, {\emph{arXiv e-prints} (2018)
  arXiv:1807.06208} [\href{https://arxiv.org/abs/1807.06208}{{\ttfamily
  1807.06208}}].

\bibitem{goodman2010}
J.~{Goodman} and J.~{Weare}, \emph{{Ensemble samplers with affine invariance}},
  \href{https://doi.org/10.2140/camcos.2010.5.65}{\emph{Communications in
  Applied Mathematics and Computational Science, Vol.~5, No.~1, p.~65-80, 2010}
  {\bfseries 5} (2010) 65}.

\bibitem{emcee2013}
D.~{Foreman-Mackey}, D.~W. {Hogg}, D.~{Lang} and J.~{Goodman}, \emph{{emcee:
  The MCMC Hammer}}, \href{https://doi.org/10.1086/670067}{\emph{\pasp}
  {\bfseries 125} (2013) 306}
  [\href{https://arxiv.org/abs/1202.3665}{{\ttfamily 1202.3665}}].

\bibitem{jeffreys1961}
H.~Jeffreys, \emph{Theory of probability}. Oxford University Press, Oxford,
  England, third~ed., 1961.

\bibitem{weeks2004}
J.~R. {Weeks}, \emph{{Maxwell's Multipole Vectors and the CMB}}, {\emph{arXiv
  e-prints} (2004) astro}
  [\href{https://arxiv.org/abs/astro-ph/0412231}{{\ttfamily
  astro-ph/0412231}}].

\bibitem{copi2004}
C.~J. {Copi}, D.~{Huterer} and G.~D. {Starkman}, \emph{{Multipole vectors: A
  new representation of the CMB sky and evidence for statistical anisotropy or
  non-Gaussianity at $2\leq \ell \leq 8$}},
  \href{https://doi.org/10.1103/PhysRevD.70.043515}{\emph{\prd} {\bfseries 70}
  (2004) 043515} [\href{https://arxiv.org/abs/astro-ph/0310511}{{\ttfamily
  astro-ph/0310511}}].

\bibitem{polastri2015}
L.~{Polastri}, A.~{Gruppuso} and P.~{Natoli}, \emph{{CMB low multipole
  alignments in the {\ensuremath{\Lambda}}CDM and dipolar models}},
  \href{https://doi.org/10.1088/1475-7516/2015/04/018}{\emph{Journal of
  Cosmology and Astro-Particle Physics} {\bfseries 2015} (2015) 018}
  [\href{https://arxiv.org/abs/1503.01611}{{\ttfamily 1503.01611}}].

\end{thebibliography}\endgroup

\end{document}